\documentclass[journal]{IEEEtran}
\usepackage{changes}

\usepackage{cite}
\usepackage{color}
\usepackage{xcolor}
\usepackage{graphicx}
\graphicspath{{Figs/}{../pdf/}{../jpeg/}}
\DeclareGraphicsExtensions{.pdf,.jpeg,.png}
\usepackage{lipsum}
\usepackage{amsmath}
\usepackage{amssymb}
\usepackage{mathrsfs}
\usepackage{algorithm}
\usepackage{array}
\usepackage{fixltx2e}
\usepackage{stfloats}
\usepackage{booktabs}
\usepackage{enumerate}
\usepackage{setspace}
\usepackage{enumitem}

\usepackage{tikz}

\usepackage[hidelinks]{hyperref}

\usepackage{amsmath}
\usepackage{amsthm}

\definecolor{CBLUE}{RGB}{0,112,192}
\definecolor{CRED}{RGB}{192,0,0}
\definecolor{CYELLOW}{RGB}{234,150,13}
\definecolor{CGREEN}{RGB}{119,172,48}
\definecolor{CGRAY}{RGB}{200,200,200}
\definecolor{CBLACK}{RGB}{80,80,80}
\definecolor{CPURPLE}{RGB}{171,68,193}

\definecolor{LBLUE}{RGB}{106,192,255}
\definecolor{LYELLOW}{RGB}{249,206,134}

\newcommand\crule[3][black]{\textcolor{#1}{\rule{#2}{#3}}}

\usepackage[normalem]{ulem}
\newcommand\rout{\bgroup\markoverwith{\textcolor{red}{/}}\ULon} 

\usepackage{todonotes}

\begin{document}

\title{Decentralized Data-Enabled Predictive Control for Power System Oscillation Damping}

\author{Linbin Huang, Jeremy Coulson, John Lygeros, and Florian D{\"o}rfler
\thanks{The authors are with the Department of Information Technology and Electrical Engineering at ETH Z{\"u}rich, Switzerland. (Emails: \text{linhuang@ethz.ch}, \text{jcoulson@control.ee.ethz.ch}, \text{jlygeros@ethz.ch}, \text{dorfler@ethz.ch})}
\thanks{This research was supported by ETH Zurich Funds.}
}


\maketitle


\begin{abstract}
We employ a novel data-enabled predictive control (DeePC) algorithm in voltage source converter (VSC) based high-voltage DC (HVDC) stations to perform safe and optimal wide-area control for power system oscillation damping. Conventional optimal wide-area control is model-based. However, in practice detailed and accurate parametric power system models are rarely available. In contrast, the DeePC algorithm uses only input/output data measured from the unknown system to predict the future trajectories and calculate the optimal control policy. We showcase that the DeePC algorithm can effectively attenuate inter-area oscillations even in the presence of measurement noise, communication delays, nonlinear loads, and uncertain load fluctuations. We investigate the performance under different matrix structures  as data-driven predictors. Furthermore, we derive a novel Min-Max DeePC algorithm to be applied independently in multiple VSC-HVDC stations to mitigate inter-area oscillations, which enables decentralized and robust optimal wide-area control. Further, we discuss how to relieve the computational burden of the Min-Max DeePC by reducing the dimension of prediction uncertainty and how to leverage disturbance feedback to reduce the conservativeness of robustification. We illustrate our results with high-fidelity, nonlinear, and noisy simulations of a four-area test system.
\end{abstract}

\vspace{-1mm}


\vspace{-2mm}

\section{Introduction}

Low-frequency inter-area oscillations prevailing in bulk power systems are generally caused by the fast exciters of synchronous generators (SGs) and long transmission lines \cite{kundur1994power}. Restraining such oscillations is essential for the secure operations of power systems. A standard solution is to implement power system stabilizers (PSSs) in the excitation system of SGs. There have been abundant works on the design of PSSs, e.g., control structure design \cite{kamwa2005ieee}, optimal control design \cite{abdel2003optimal,jain2015model,wu2015input} and decentralized design \cite{wu2015input,kamwa2001wide,nabavi2015distributed}. The appropriate placement of PSSs can be obtained from participation factors (by using Prony method, etc.) or transfer function residues \cite{breulmann2000analysis}.

Another popular solution is to utilize the high controllability and flexibility of high-voltage DC (HVDC) stations to mitigate low-frequency oscillations \cite{azad2015decentralized,summers2015submodularity,bjork2018fundamental,huang2018damping,perez2018virtual}. Unlike SGs, HVDC stations are three-phase power converters which have no rotational part and thus enable fast voltage magnitude and phase control in power grids.
It has been shown in \cite{bjork2018fundamental,huang2018damping,perez2018virtual} that with proper control design, the voltage source converter (VSC) based HVDC station can effectively mitigate low-frequency oscillations. Moreover, with wide area measurement systems (WAMS), optimal control can be performed in VSC-HVDC stations by employing model predictive control (MPC) {or linear quadratic Gaussian (LQG) control} to stabilize the system \cite{fuchs2013stabilization,azad2013damping,yu1970application}.
In fact, the application of WAMS greatly facilitates system identification based on the Phasor Measurement Units (PMUs) data and the subsequent control design \cite{chakrabortty2012wide}.
{It was shown in \cite{fuchs2013stabilization} that the MPC-based damping controller with wide-area measurements has superior performance than local damping controllers. It was reported in \cite{azad2013damping} that an MPC-based damping controller has faster damping than an LQG-based damping controller thanks to the time-varying gains obtained from online optimization. Moreover, input and output constraints can be conveniently included in MPC-based damping controllers.}
However, an accurate and detailed model of the system is needed for the controller design or prediction of the future behaviours, which may result in inferior performance under model mismatch or uncertainties.

Normally, the uncertainties in the system are handled using robust or adaptive methods. For example, the value set approach was used in \cite{zhou2017large} to perform robust stability analysis and parameter design in large power systems. A robust design of multi-machine PSSs based on simulated annealing optimization technique was presented in \cite{abido2000robust}. However, these methods are still model-based and thereby result in complicated design and complex controllers. We note that although model-based design in theory provides an optimal solution for the oscillation events, optimality and robustness can rarely be achieved in practice because (i) the true parameters of the devices (e.g., HVDC stations and SGs) are hard to obtain due to dependency on operating conditions and parameter uncertainty; (ii) the control algorithms of the devices designed by their manufacturers are usually unknown from the system operator's point of view; (iii) the grid model is ever-changing and thereby hard to obtain due to different operation modes, uncertainties, and relaying; {(iv) conventional model-based control design and control tuning require significant effort by commissioning engineers}. To tackle such challenges, recent control approaches entirely circumvent these model-based methods in favor of data-driven approaches \cite{lewis2012reinforcement,dean2017sample,de2019persistency}.

In our previous work \cite{coulson2019data,huang2019data,coulson2019regularized} we have developed a novel {\bf D}ata-{\bf e}nabl{\bf e}d {\bf P}redictive {\bf C}ontrol (DeePC) algorithm and applied it to a VSC-HVDC station to perform safe and optimal control, which uses local measurements to effectively eliminate the oscillations in a two-area system. The DeePC algorithm needs only input/output measurements from the unknown system to predict the future trajectory and uses the real-time feedback to drive the unknown system along a desired optimal trajectory \cite{coulson2019data}. {The stability of DeePC was investigated in \cite{berberich2019data} which showed that the regularizations and terminal equality constraints can be used to provide stability guarantees even in the presence of measurement noise and corrupted data.} The utility of DeePC for grid-connected converters has been show-cased in \cite{huang2019data}.

Rather than a parametric system representation, the DeePC algorithm proposed in \cite{coulson2019data} relies on a behavioral system approach which describes the input/output behaviour of the system through the subspace of the signal space wherein trajectories of the system live \cite{markovsky2006exact,willems2005note,markovsky2008data}. This signal space of trajectories is spanned by the columns of a data Hankel matrix which results in a non-parametric and data-centric perspective on dynamical control systems.

The original contributions of this paper are as follows. We
 apply the DeePC algorithm in multiple VSC-HVDC stations to perform optimal wide-area control for power system oscillation damping. 
In a first step, DeePC is deployed in a large-scale case study as a centralized controller which provides optimal control signals for multiple VSC-HVDC stations. {We note that due to the data-centric system representation, the DeePC algorithm implicitly takes into account the impact of unknown (albeit constant) communication delays, as long as an upper bound is known, and we can collect enough data to construct Hankel matrices of appropriate dimension.}
We test the performance of the DeePC algorithm under various system settings and compare it to certainty-equivalence MPC relying on a nominal model. It is shown that DeePC achieves better performance even in the presence of noisy measurements and system nonlinearity. Furthermore, we compare the performance of the DeePC algorithm when using a Hankel or Page matrix structure. The Page matrix is also known as a predictive time series matrix \cite{agarwal2018model,damen1982approximate} and leads to superior performance. We also investigate how the performance can be further improved by employing a denoising process on the Page matrix based on singular value decomposition (SVD).

We then develop a Min-Max DeePC algorithm which enables decentralized, robust, and optimal wide-area control and discuss how to reduce the computational burden of the Min-Max DeePC and to achieve real time implementation. Moreover, we develop a disturbance-feedback (DF) Min-Max DeePC algorithm to reduce the conservativeness of robustification and to leverage disturbance feedback. All of our results are illustrated with high-fidelity nonlinear simulations.

The rest of this paper is organized as follows: in Section II we give a brief review on the DeePC algorithm. Section III applies DeePC in a four-area test systems to perform optimal wide-area control. In Section IV we present the Min-Max DeePC and discuss how to reduce the computational burden. Section V applies the Min-Max DeePC in the four-area test system to perform robust and optimal wide-area control in a decentralized way. We conclude the paper in Section VI.

\section{Data-Enabled Predictive Control}


\subsection{Preliminaries and Notation}

Consider an $n{\rm th}$-order minimal discrete-time linear time-invariant (LTI) system
\begin{equation}
\left\{ \begin{array}{l}
{x_{t + 1}} = A{x_t} + B{u_t}\\
{y_t} = C{x_t} + D{u_t}
\end{array} \right.\,,		\label{eq:ABCD}
\end{equation}
where $A \in \mathbb{R}^{n \times n}$, $B \in \mathbb{R}^{n \times m}$, $C \in \mathbb{R}^{p \times n}$, $D \in \mathbb{R}^{p \times m}$, $x_t \in \mathbb{R}^n$ is the state of the system, $u_{t} \in \mathbb{R}^m$ is the input vector, and $y_{t} \in \mathbb{R}^p$ is the output vector at times $t \in \{0,1,2,\dots\}$, where $t$ takes value on the discrete-time axis $\mathbb{Z}_{ \ge 0}$. Let $u_{i,t}$ be the $i{\rm th}$ element of $u_t$ and $y_{i,t}$ the $i{\rm th}$ element of $y_t$.

The \textit{lag} of the system in (\ref{eq:ABCD}) is defined by the smallest integer $\ell \in \mathbb{Z}_{ \ge 0}$ such that the observability matrix
\vspace{-0.5mm}
\begin{equation*}
\mathscr{O}_{\ell}(A,C) := {\rm{col}}(C,CA,...,CA^{\ell-1})
\vspace{-0.5mm}
\end{equation*}
has rank $n$, i.e., the state can be reconstructed from $\ell$ measurements. Here ${\rm{col}}(a_0,a_1,...,a_i):=[a_0^{\top}\; a_1^{\top}\; \cdots \;a_i^{\top}]^{\top}$.

Let $u = {\rm{col}}(u_0,u_1,u_2,...)$ and $y = {\rm{col}}(y_0,y_1,y_2,...)$ be the input and output trajectories with dimensions inferred from the context.
Let $L,T \in \mathbb{Z}_{ \ge 0}$. The trajectory $u \in \mathbb{R}^{mT}$ is \textit{persistently exciting of order L} if the block Hankel matrix (of \textit{depth} $L$)\footnote{Unlike the definition in linear algebra studies which requires Hankel matrices to be square, here we follow the convention of behavioral systems theory and subspace identification \cite{willems2005note,markovsky2006exact} and allow general dimensions.}
\begin{equation}
\mathscr{H}_L(u) := \left[ {\begin{array}{*{20}{c}}
	{{u_0}}&{{u_1}}& \cdots &{{u_{T - L}}}\\
	{{u_1}}&{{u_2}}& \cdots &{{u_{T - L + 1}}}\\
	\vdots & \vdots & \ddots & \vdots \\
	{{u_{L-1}}}&{{u_{L}}}& \cdots &{{u_{T-1}}}
	\end{array}} \right] 		
\label{eq:Hankel_L}
\end{equation}
is of full row rank, i.e., the signal $u$ is sufficiently rich and sufficiently long. Note that a necessary condition for persistency of excitation is $T\ge (m+1)L-1$ \cite{willems2005note,coulson2019data}.

Consider $T_{\rm ini},N,T \in \mathbb{Z}_{ \ge 0}$ such that $T \ge (m+1)(T_{\rm ini}+N+n)-1$, a length-$T$ input trajectory $u^{\rm{d}} \in \mathbb{R}^{mT}$ that is persistently exciting of order $T_{\rm ini} + N + n$ and the corresponding length-$T$ output trajectory $y^{\rm{d}} \in \mathbb{R}^{pT}$ measured from \eqref{eq:ABCD}. The superscript d is used to indicate that $u^{\rm d}$ and $y^{\rm d}$ are sequences of input/output data samples measured from the system \eqref{eq:ABCD}. Here we assume that the state-space matrices $A$, $B$, $C$ and $D$ are unknown. We use $u^{\rm{d}}$ and $y^{\rm{d}}$ to construct the Hankel matrices $\mathscr{H}_{T_{\rm ini}+N}(u^{\rm{d}})$ and $\mathscr{H}_{T_{\rm ini}+N}(y^{\rm{d}})$, which are further partitioned into two parts
\begin{equation}
\left[ {\begin{array}{*{20}{c}}
	{{U_{\rm P}}}\\
	{{U_{\rm F}}}
	\end{array}} \right] := \mathscr{H}_{T_{\rm ini}+N}(u^{\rm{d}})\,,\;\;\left[ {\begin{array}{*{20}{c}}
	{{Y_{\rm P}}}\\
	{{Y_{\rm F}}}
	\end{array}} \right] := \mathscr{H}_{T_{\rm ini}+N}(y^{\rm{d}})\,,		\label{eq:partition_Huy}
\end{equation}
where $U_{\rm P} \in \mathbb{R}^{mT_{\rm ini} \times H_c}$, $U_{\rm F} \in \mathbb{R}^{mN \times H_c}$, $Y_{\rm P} \in \mathbb{R}^{pT_{\rm ini} \times H_c}$, $Y_{\rm F} \in \mathbb{R}^{pN \times H_c}$, and $H_c = T-T_{\rm ini}-N+1$. We remark that the above Hankel matrices are constructed from the input/output trajectories $u^{\rm d}$ and $y^{\rm d}$ which are measured offline before the DeePC algorithm is applied. During this data-collection period, the control inputs can be white noise signals, to make $u^{\rm d}$ persistently exciting of order $T_{\rm ini}+N+n$.
In the sequel, the data in the partition with subscript P (for ``past'') will be used to implicitly estimate the initial condition of the system, whereas the data with subscript F will be used to predict the ``future'' trajectories. Here $T_{\rm ini}$ is the length of an initial trajectory and $N$ is the length of a predicted trajectory starting from the initial trajectory (i.e., we predict forward $N$ steps).

According to the Fundamental Lemma in \cite{willems2005note}, ${\rm{col}}(u_{\rm ini}, y_{\rm ini}, u, y)$ is a trajectory of (\ref{eq:ABCD}) if and only if there exists $g \in \mathbb{R}^{H_c}$ such that
\begin{equation}
\left[ {\begin{array}{*{20}{c}}
	{{U_{\rm P}}}\\
	{{Y_{\rm P}}}\\
	{{U_{\rm F}}}\\
	{{Y_{\rm F}}}
	\end{array}} \right]g = \left[ {\begin{array}{*{20}{c}}
	{{u_{\rm ini}}}\\
	{{y_{\rm ini}}}\\
	u\\
	y
	\end{array}} \right]\,,		\label{eq:Hankel_g}
\end{equation}
where $u_{\rm ini} \in \mathbb R^{mT_{\rm ini}}$, $y_{\rm ini} \in \mathbb R^{pT_{\rm ini}}$, $u \in \mathbb R^{mN}$, and $y \in \mathbb R^{pN}$.
The trajectory ${\rm{col}}(u_{\rm ini},y_{\rm ini})$ (of length $T_{\rm ini}$) can be thought of as setting the initial condition for the future trajectory ${\rm{col}}(u,y)$ (of length $N$), and  ${\rm{col}}(u_{\rm ini},y_{\rm ini},u,y)$ is the entire trajectory.

If $T_{\rm ini} \ge \ell$, the future output trajectory $y$ is uniquely determined through (\ref{eq:Hankel_g}) for every given input trajectory $u$ \cite{markovsky2008data}. {A recent result in \cite{markovskyidentifiability} extends the Fundamental Lemma to consider mosaic Hankel, Page, and trajectory matrix structures, that only requires ${\rm rank}(\mathscr{H}_{T_{\rm ini}+N}(u,y)) = m(T_{\rm ini}+N)+n$.}

In a data-driven setting, $\ell$ and $n$ are not known, and we can use an upper bound on them instead (see Section~\ref{Section_Centralized_Wide-Area_Control} for the parameter tuning of DeePC). Also, one should try to make the bound tight for computational and overfitting reasons.

\subsection{Review of DeePC}

The DeePC algorithm \cite{coulson2019data} uses input/output data collected from the unknown system to predict the future behaviour and perform optimal and safe control, thereby avoiding a parametric system representation. After using the input/output trajectory ${\rm{col}}(u^{\rm{d}},y^{\rm{d}})$ ($u^{\rm{d}} \in \mathbb{R}^{mT}$ and $y^{\rm{d}} \in \mathbb{R}^{pT}$) to construct the Hankel matrices in (\ref{eq:partition_Huy}), DeePC solves the following optimization problem to get the optimal future control inputs
\begin{equation}
\begin{array}{l}
\mathop {{\rm{min}}}\limits_{g,\sigma_y,u \in \mathcal U, y \in \mathcal Y} \;\;{\left\| u \right\|_R^2} + {\left\| {y - r} \right\|_Q^2} + {\lambda _g}{\left\| g \right\|_2^2} + {\lambda _y}{\left\| \sigma_y \right\|_2^2}\\
{\rm s.t.}\;\;\left[ {\begin{array}{*{20}{c}}
	{{U_{\rm P}}}\\
	{{Y_{\rm P}}}\\
	{{U_{\rm F}}}\\
	{{Y_{\rm F}}}
	\end{array}} \right]g = \left[ {\begin{array}{*{20}{c}}
	{{u_{\rm ini}}}\\
	{{y_{\rm ini}}}\\
	u\\
	y
	\end{array}} \right] + \left[ {\begin{array}{*{20}{c}}
	0\\
	\sigma_y\\
	0\\
	0
	\end{array}} \right]\,,		
\label{eq:DeePC}
\end{array}
\end{equation}
where $\mathcal U \subseteq \mathbb{R}^{mN}$ and $\mathcal Y \subseteq \mathbb{R}^{pN}$ are the input and output constraint sets, $R \in \mathbb{R}^{mN \times mN}$ is the control cost matrix (positive definite), $Q \in \mathbb{R}^{pN \times pN}$ is the output cost matrix (positive semidefinite), $\sigma_y \in \mathbb{R}^{pT_{\rm ini}}$ is an auxiliary slack variable to ensure feasibility of the initial condition equality constraint, $\lambda_g,\lambda_y \in \mathbb{R}_{ \ge 0}$ are regularization parameters (we choose $\lambda_y$ sufficiently large such that the implicit initial condition estimation is accurate), $r \in \mathbb{R}^{pN}$ is the reference trajectory for the outputs, $N$ is the prediction horizon, ${\rm{col}}(u_{\rm ini},y_{\rm ini})$ consists of the most recent input/output trajectory of (\ref{eq:ABCD}) of length $T_{\rm ini}$, and ${\left\| a \right\|_X^2}$ denotes the quadratic form $a^\top Xa$.

A quadratic two-norm penalty on $g$ is included in the cost function as a regularization term to avoid overfitting in case of noisy data samples. In fact, when stochastic disturbances affect the output measurements, a two-norm regularization of $g$ coincides with distributional two-norm robustness in the trajectory space \cite{coulson2019regularized}. {It was further shown in \cite{huang2020quadratic} and \cite{huang2021robust} that including such a quadratic regularization of $g$ in \eqref{eq:DeePC} is equivalent to solving a min-max optimization problem that minimizes the worst-case performance for bounded disturbance sets on the input/output data.}

DeePC involves solving the optimization problem (\ref{eq:DeePC}) in a receding horizon manner, that is, after calculating the optimal control sequence $u^\star$, we apply $(u_t,...,u_{t+k-1}) = (u_0^{\star},...,u_{k-1}^{\star})$ to the system for some $k \le N$ time steps, update ${\rm{col}}(u_{\rm ini},y_{\rm ini})$ to the most recent input/output measurements, and then set $t$ to $t+k$ for the DeePC algorithm. {We refer to $k$ as the control horizon in this paper.}

{Terminal constraints can also be included in \eqref{eq:DeePC} for stability \cite{berberich2019data}. However, we refrain from doing so because we consistently observed similar performance with and without the terminal constraints. Also, the optimization problem can be solved faster without the terminal constraints.}

Earlier work \cite{huang2019data} has shown how DeePC is related to certainty-equivalence MPC, i.e., based on a nominal model. To be specific, an $N$-step auto-regressive model with extra input (ARX) of the system can be identified using a least-square multi-step prediction error method (PEM) as \cite[Lemma 3.1]{huang2019data}
\begin{equation}
y=Y_{\rm F}\left[\begin{array}{c}{U_{\rm P}} \\ {Y_{\rm P}} \\ {U_{\rm F}}\end{array}\right]^{+}\left[\begin{array}{c}{u_{\mathrm{ini}}} \\ {y_{\mathrm{ini}}} \\ {u}\end{array}\right]\,,
\label{eq:ARMA}
\end{equation}
where the superscript $+$ denotes the pseudoinverse operator.

Then, the certainty-equivalence PEM-MPC solves the following optimization problem in a receding horizon manner
\begin{equation}
\begin{array}{l}
\mathop {\min }\limits_{u \in \mathcal U,y \in \mathcal Y} \;\;{\left\| u \right\|_R^2} + {\left\| {y - r} \right\|_Q^2}\\
\;\;\;{\rm s.t.}\;\;\;\;\;\;\;\eqref{eq:ARMA}\,.	
\end{array}	
\label{eq:MPC}
\end{equation}

In fact, obtaining the ARX model from the Hankel matrices in \eqref{eq:ARMA} coincides with solving \eqref{eq:Hankel_g} for $y = Y_{\rm F}g$ and
\begin{equation}
g=\left[\begin{array}{c}{U_{\rm P}} \\ {Y_{\rm P}} \\ {U_{\rm F}}\end{array}\right]^{+}\left[\begin{array}{c}{u_{\mathrm{ini}}} \\ {y_{\mathrm{ini}}} \\ {u}\end{array}\right]\,,
\label{eq:PEM_g}
\end{equation}
which is the least-norm solution that satisfies the constraints in \eqref{eq:DeePC} when $\sigma_y=0$; in this sense, DeePC provides more flexibility in representing the unknown system \cite[Lemma 3.2]{huang2019data} rather than using the particular identified model \eqref{eq:ARMA}. We will compare the performance of DeePC and PEM-MPC in Section~\ref{Section_Centralized_Wide-Area_Control}; {we also refer to \cite{dorfler2021bridging} for a formal comparison of the two methods.}

\subsection{DeePC with Page Matrix}
\label{subsec: DeePC with Page Matrix}

As outlined above, previous work on the DeePC algorithm relies on arranging the input/output data, i.e., $u^{\rm d}$ and $y^{\rm d}$, into block Hankel matrices for predicting the future system behavior. Here we also explore the alternative arrangement of the data into block (Chinese) Page matrices \cite{agarwal2018model,damen1982approximate} of the following form (assuming that $T$ is a multiple of $L$)
\begin{equation}
\mathscr{P}_L(u^{\rm d}) := \left[ {\begin{array}{*{20}{c}}
	{{u_0}}&{{u_L}}& \cdots &{{u_{T - L}}}\\
	{{u_1}}&{{u_{L+1}}}& \cdots &{{u_{T - L + 1}}}\\
	\vdots & \vdots & \ddots & \vdots \\
	{{u_{L-1}}}&{{u_{2L-1}}}& \cdots &{{u_{T-1}}}
	\end{array}} \right] \,.
\label{eq:Page_L}
\end{equation}
Similar to the partitioning in \eqref{eq:partition_Huy}, we obtain $U_{\rm P}$, $U_{\rm F}$, $Y_{\rm P}$, and $Y_{\rm F}$ from $\mathscr{P}_{T_{\rm ini}+N}(u^{\rm d})$ and $\mathscr{P}_{T_{\rm ini}+N}(y^{\rm d})$ and use them for predicting the system as in \eqref{eq:Hankel_g} and \eqref{eq:DeePC}, replacing all Hankel matrices used in DeePC by Page matrices.

Both Hankel and Page matrices serve as data-driven predictors, but the latter has a few advantages, as pointed out in \cite{agarwal2018model,damen1982approximate}.
The key difference between the two is that none of the entries in the Page matrix are repeated. This has both advantages and disadvantages. The main disadvantage is that more data is needed to construct the matrix. On the other hand, if the measurements are subject to noise, the entries of the Page matrix are statistically independent. As a consequence, the measurement noise in the output signals can be filtered by performing singular value decomposition (SVD) on the Page matrices and then truncating the small singular values, without breaking the structure of the data matrices \cite{damen1982approximate}.

To be specific, we assume that noise is uncorrelated for different measurements and de-noise them one-by-one: for the $i{\rm th}$ output, we denote its trajectory of length $T$ as $y^{\rm d}_{i,\cdot} = {\rm col}(y^{\rm d}_{i,0},y^{\rm d}_{i,1},...,y^{\rm d}_{i,T-1})$, and perform SVD on $\mathscr{P}_{T_{\rm ini}+N}(y^{\rm d}_{i,\cdot})$:
\begin{equation}\label{eq:SVD}
\mathscr{P}_{T_{\rm ini}+N}(y^{\rm d}_{i,\cdot}) = U\Sigma V^\top \,,
\end{equation}
where $\Sigma \in \mathbb{R}^{(T_{\rm ini}+N) \times H_c}$ is a rectangular diagonal matrix of singular values, and $U \in \mathbb{R}^{(T_{\rm ini}+N) \times (T_{\rm ini}+N)}$ and $V \in \mathbb{R}^{H_c \times H_c}$ are unitary matrices.
Next, we replace by zeros the singular values in $\Sigma$ that are smaller than a noise dependent threshold $\sigma_0$. This is motivated by the results in the identification and low-rank approximation literature \cite{damen1982approximate,chatterjee2015matrix} that suggest that removing small singular values is equivalent to filtering out noise.

Let $\Sigma'$ be the new singular value matrix after the above noise-filtering process. Based on $\Sigma'$, the noise-filtered Page matrix of the $i{\rm th}$ output can be constructed as
\begin{equation}\label{eq:SVD'}
\mathscr{P}'_{T_{\rm ini}+N}(y^{\rm d}_{i,\cdot}) = U\Sigma' V^\top \,.
\end{equation}

After filtering the $p$ outputs one-by-one, the $p$ noise-filtered Page matrices $\mathscr{P}'_{T_{\rm ini}+N}(y^{\rm d}_{i,\cdot})$ ($i \in \{1,2,...,p\}$) can be stacked to obtain the noise-filtered block Page matrix as
\begin{equation}\label{eq:YPYF'}
\left[ {\begin{array}{*{20}{c}}	{{Y'_{\rm P}}}\\	{{Y'_{\rm F}}} \end{array}} \right] = \sum\limits_{i=1}^{p} { \mathscr{P}'_{T_{\rm ini}+N}(y^{\rm d}_{i,\cdot}) \otimes e_i^p } \,,
\end{equation}
where $\otimes$ denotes the Kronecker product, $e_i^p \in \mathbb{R}^p$ is a vector with entry 1 at position $i$ and 0 at all other positions. Note that $Y_{\rm P} = Y'_{\rm P}$ and $Y_{\rm F} = Y'_{\rm F}$ if setting $\sigma_0 = 0$, i.e., without noise filtering. We will show that the performance of the DeePC algorithm can be significantly improved by employing (i) the Page matrix structure and (ii) the noise filtering based on singular-value thresholding. Observe that a similar de-noising of Hankel matrices leads to filtered matrices, which have no Hankel structure and thus cannot serve as predictors for LTI systems as in \eqref{eq:Hankel_g}; indeed, our results reported below suggest that this tends to lead to poor performance (Section~\ref{subsec: Comparison of Hankel Matrix and Page Matrix}).

In addition to Hankel and Page matrix structures, it is also possible to use other matrix structures as predictors, e.g., a concatenation of many thin Hankel matrices \cite{van2020willems} allowing for multiple short experiments rather than a single long one.

\section{Centralized Wide-Area Control}
\label{Section_Centralized_Wide-Area_Control}

In this section we apply the DeePC algorithm to VSC-HVDC stations, to perform centralized optimal wide-area control so as to mitigate low-frequency oscillations. Note that compared to \cite{huang2019data}, in what follows we consider a much more realistic, large-scale, and challenging system setup. Particularly, the DeePC algorithm will be employed in a VSC-HVDC link (rather than a single station) considering the dynamic interaction between two VSC-HVDC stations.

\begin{figure*}[!t]
	\centering
	\includegraphics[width=6.3in]{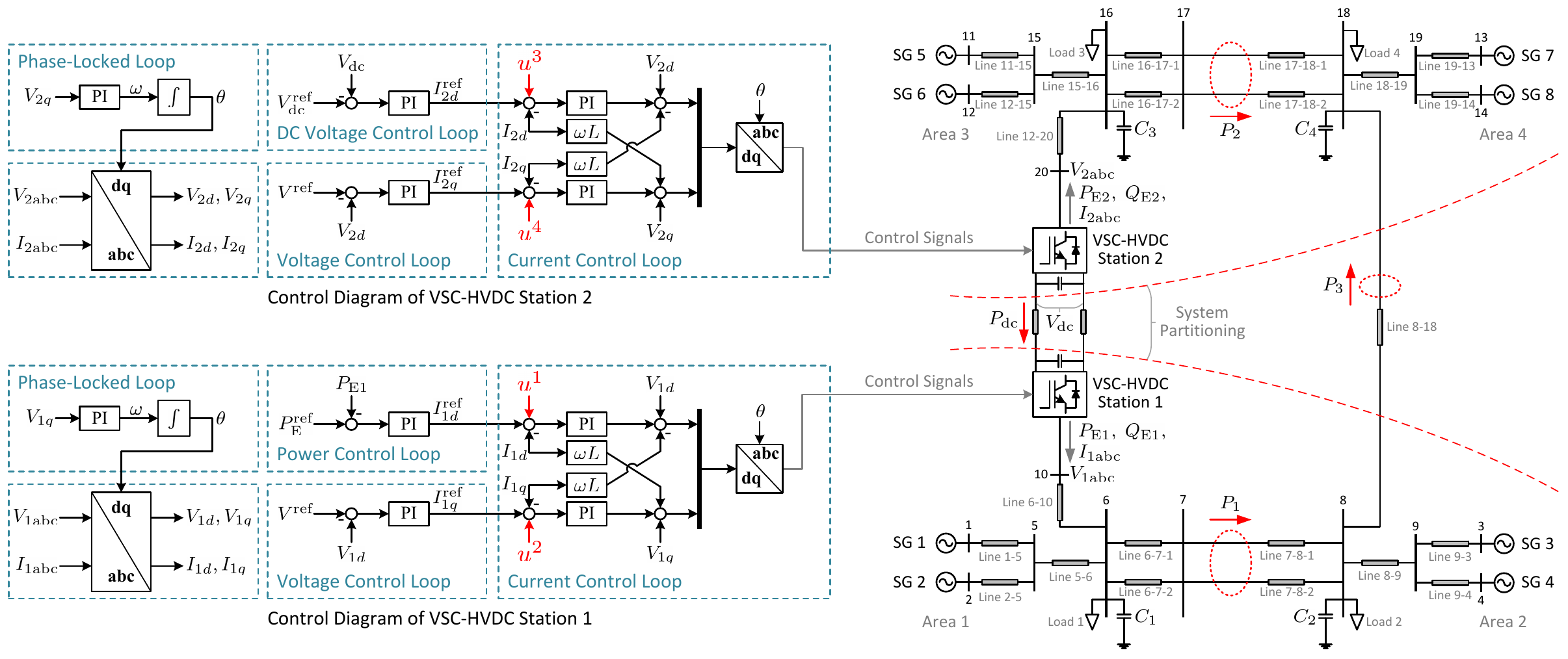}
	\vspace{-4mm}
	\caption{One-line diagram of a four-area test system with integration of an HVDC link.}
	\vspace{-1mm}
	\label{Fig_four_area_syst}
\end{figure*}

\subsection{Descriptions of a Four-area Test System}

Though the approach is general, to illustrate the point we consider a four-area test system with integration of an HVDC link in Fig.~\ref{Fig_four_area_syst}. The system has $n=208$ states. The main parameters of this system are given in Table \ref{table:sys_parameters} in Appendix \ref{Appendix: System parameters}. The four-area system has weakly-damped inter-area oscillations due to the fast exciters in SGs and long transmission lines.

The VSC-HVDC station 1 performs active power control in order to regulate the power flow of the DC link, and the VSC-HVDC station 2 performs DC voltage control for the HVDC link. Both of the VSC-HVDC stations apply phase-locked loops to synchronize with the AC grid and voltage control loops to regulate their terminal voltage. Generally, the conventional control structures of VSC-HVDC stations as shown in Fig.~\ref{Fig_four_area_syst} do not have enough control freedom to achieve the functionality of oscillation damping, and thus auxiliary control is needed \cite{azad2015decentralized,bjork2018fundamental}. In fact, in addition to using a VSC-HVDC link for oscillation damping, our algorithms in this paper also have the potential to be applied in the excitation systems of SGs (by choosing different control inputs) and achieve a similar functionality to conventional power system stabilizers, but in a model-free and data-driven manner.

Note that the four-area system in this paper is an extension of the two-area benchmark model for power system stability studies \cite{canizares2016benchmark} (by replicating the model twice) in order to integrate a VSC-HVDC link. As will be shown below, this four-area system has sustained low-frequency oscillations if auxiliary damping control is not applied, that is, the dominant poles are close to the imaginary axis. In this paper we do not provide modal analysis for the system since we focus on model-free approaches to eliminate power system oscillations.

\subsection{Centralized Wide-area Control Using DeePC}
\label{subsec:Centralized Wide-area Control}

\begin{figure}[!t]
	\centering
	\includegraphics[width=2.8in]{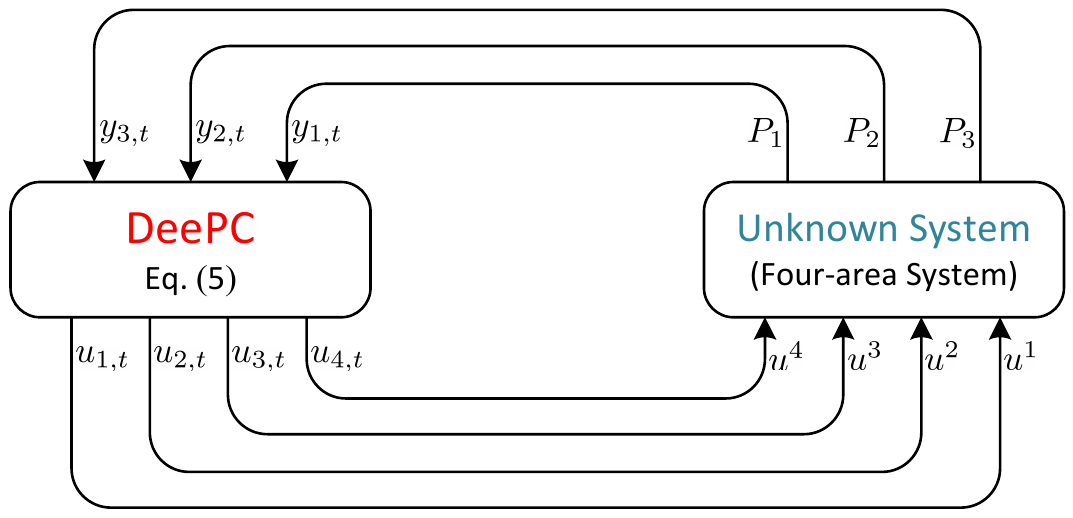}
	\vspace{-3mm}
	\caption{Centralized wide-area control based on DeePC.}
	\vspace{-1mm}
	\label{Fig_Centralized_DeePC}
\end{figure}

We present now a centralized wide-area control based on DeePC as shown in Fig.~\ref{Fig_Centralized_DeePC}. The controller collects the wide-area measurements of $P_1$, $P_2$ and $P_3$ (which are respectively the interface power flows from Bus 7 to Bus 8, from Bus 17 to Bus 18, and from Bus 8 to Bus 18 as labeled in Fig.~\ref{Fig_four_area_syst}), and then distributes the optimal control inputs to the two VSC-HVDC stations through $u^1$, $u^2$, $u^3$ and $u^4$ as displayed in Fig.~\ref{Fig_four_area_syst}. {These control inputs (unlike active/reactive power) merely affect transient performance with a high bandwidth and have no impact on the steady state, that is determined by the PI regulators in the outer loops. The selection of these input/output signals is similar to conventional model-based damping control schemes implemented in HVDC stations, which have been proven to be effective in damping low-frequency oscillations \cite{azad2013damping}; this is sufficient for applying our method to the system, as it suggests that all the modes we would like to regulate are indeed controllable from the chosen inputs.}

{Unknown (albeit constant) communication and measurement delays can be considered as part of the unknown system by constructing data matrices of appropriate dimensions and their impact is implicitly taken into account in the DeePC algorithm. Thanks to the robustifying regularizations, we expect DeePC to be robust to variable delays within a small range.
However, if the delays are significantly varying over time, an adaptive version of DeePC would be needed. For example, since the time delay is sometimes known (due to time-stamped measurements), one can update the data in the Hankel matrices when detecting significant changes of the delay; this topic will be investigated in future work.}

\vspace{2mm}
{\noindent \em Configuration of the DeePC Algorithm}

\begin{itemize}[leftmargin=*]
	\item The sampling time of DeePC is chosen as $0.02{\rm s}$ since we focus on low-frequency dynamics here. Notice that the sampling time of DeePC is different from that of the basic control schemes of the VSC-HVDC stations ($10{\rm kHz}$).
	\item We choose the length of the initial trajectory to be $T_{\rm ini} = 60$ and assume that it is greater than the lag of the unknown system. The prediction horizon is chosen to be $N = 120$.
	\item The parameters in the cost function are set to $R = I$, $Q = 400\times I$, $\lambda_g = 20$ and $\lambda_y = 2000$ ($I$ is the identity matrix whose dimension can be inferred from the context). The reference trajectory $r$ is set to be equal to the steady-state of $y$, which can be obtained from the power flow calculation. As an alternative, the steady-state values of $y$ can also be obtained purely from recorded data by averaging the upper and lower bounds of the measured oscillations. {Note that $y=r$ and $u=0$ is a valid steady state of the system because the control signal $u$ is added to the current reference provided by the outer loops. The integral action included in the outer loops will ensure that at steady state the current reference will converge to a value that ensures $y=r$ can be achieved with the additional input $u=0$ provided by DeePC. The same argument also extends to the PEM-MPC that serves as a basis for comparison below.}
	\item Before DeePC is activated, persistently exciting white noise signals with noise power being $10^{-4}~{\rm p.u.}$ (generated from the Band-Limited White Noise blocks in Simulink) are injected into the system through $u^1$, $u^2$, $u^3$ and $u^4$ for $30{\rm s}$ to construct the input/output Hankel matrix in \eqref{eq:partition_Huy} (with $T = 1500$). {Other types of persistently exciting input signals can be used in practice to possibly achieve better performance, e.g., a pseudorandom binary sequence \cite{elokda2019data}.}
\end{itemize}

\vspace{1mm}

\begin{figure}[!t]
	\centering
	\includegraphics[width=2.8in]{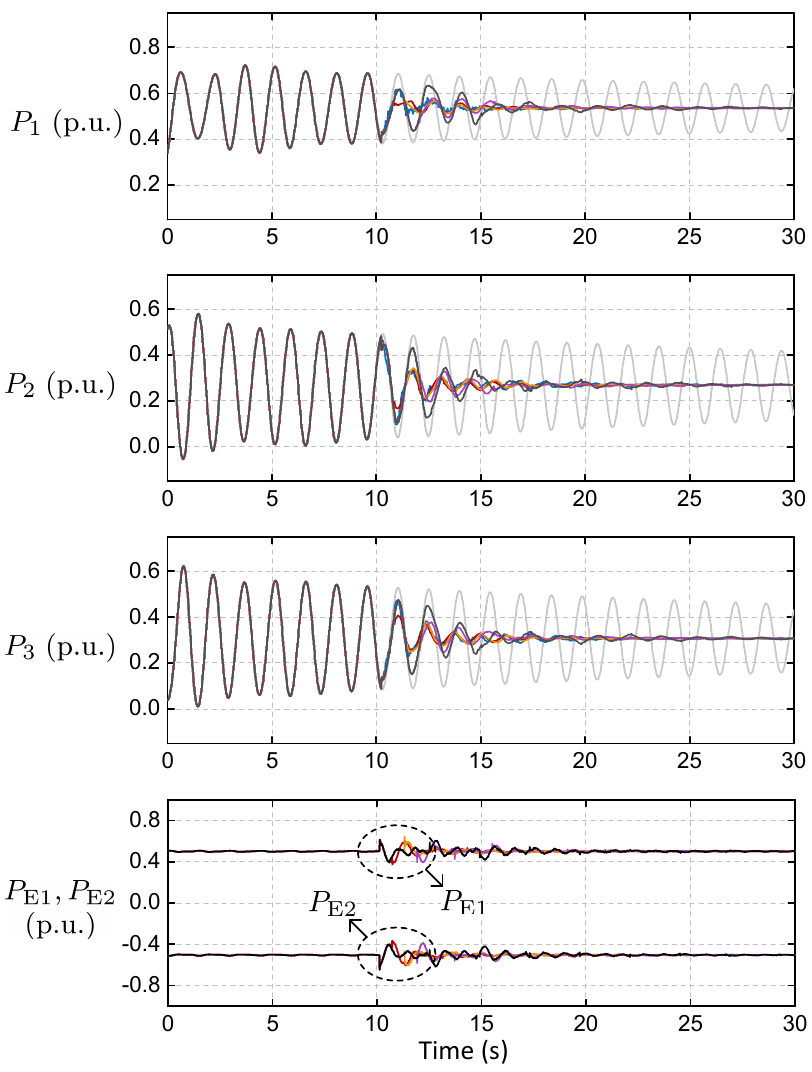}
	\vspace{-2mm}
	\caption{Time-domain responses of the four-area system. DeePC (or PEM-MPC) is activated at $t = 10{\rm s}$. {\color{CGRAY}{\bf{-----}}} without wide-area control; {\color{CBLUE}{\bf{-----}}} with PEM-MPC ($k = 60$); {\color{CRED}{\bf{-----}}} with DeePC ($k = 30$); {\color{CYELLOW}{\bf{-----}}} with DeePC ($k = 60$); {\color{CPURPLE}{\bf{-----}}} with DeePC ($k = 90$); {\color{CBLACK}{\bf{-----}}} with DeePC ($k = 120$).}
	\vspace{-1mm}
	\label{Fig_linear_cases}
\end{figure}

To illustrate the effectiveness of the DeePC algorithm, we now provide a detailed simulation study based on a nonlinear model of the four-area system given in Fig.~\ref{Fig_four_area_syst}.
As a base case, here we consider the loads to be constant power loads, and the output measurements to be noise-free (we will later consider nonlinear loads, load fluctuations and noisy measurements).
Fig.~\ref{Fig_linear_cases} displays the responses of the four-area system when the DeePC algorithm is adopted. {The starting point for our simulation is a system configuration with poorly damped inter-area modes.} We apply the first $k$ elements of the optimal control sequence to the system every time after solving \eqref{eq:DeePC}, as described in Section II B. It can be seen that DeePC effectively attenuates the inter-area oscillations after it is activated at $t = 10{\rm s}$. Moreover, the damping ratio is improved with the decrease of $k$ because of the nonlinearity of the system resulting in a prediction error. Hence, reducing $k$ introduces faster feedback and improves the real-time closed-loop performance. On the other hand, reducing $k$ increases the computational burden since the optimization problem \eqref{eq:DeePC} needs to be solved more frequently. Note that \eqref{eq:DeePC} is a standard quadratic program. This can be seen by substituting $u = U_{\rm F}g$, $y = Y_{\rm F}g$ and $\sigma_y = Y_{\rm P}g - y_{\rm ini}$ into the cost function. {The dimension of the decision variables (i.e., $g$) is $H_c$, the number of equality constraints is $mT_{\rm ini}$, and the number of inequality constraints is $2(m+p)N$ as we consider upper and lower bounds for the inputs and outputs.} Hence, the computational complexity and memory resource requirements for solving \eqref{eq:DeePC} are exactly the same as solving standard quadratic programs, which can be solved in polynomial time.
To solve the optimization problem \eqref{eq:DeePC} we use OSQP, a computationally efficient solver for quadratic programs \cite{osqp} that is also embeddable in some widely-used microcontrollers. On an Intel Core i5 7200U CPU with 8GB RAM, OSQP requires approximately $1{\rm s}$ to solve \eqref{eq:DeePC} every time in the above simulations. Therefore, by setting $k$ larger than 50 (the sampling time is $0.02{\rm s}$), DeePC can be solved in real time, even without further customization or optimization of the code.

The active power responses of the two VSC-HVDC stations are given in Fig.~\ref{Fig_linear_cases}, which shows that power fluctuations during the transient are acceptable. Such fluctuations arise due to the fact that the VSC-HVDC stations participate in the low-frequency oscillations (otherwise, their active power will remain constant). Note that by choosing a proper input constraint set $\mathcal U$, the active power fluctuations of the VSC-HVDC stations can be limited within the admissible range. One can also penalize the rates of change of the inputs ($\Delta u_i = u_i - u_{i-1}$, $i=\{1,\dots,N-1\}$) in the cost function of \eqref{eq:DeePC} to smoothen the transient responses of the stations.

Fig.~\ref{Fig_linear_cases} plots the system responses when certainty-equivalence PEM-MPC is applied in the wide-area controller, with the same data, $Q$, and $R$ as DeePC. It can be seen that in this case PEM-MPC effectively eliminates the inter-area oscillations as well, with the damping performance slightly worse than the DeePC algorithm (both with $k = 60$).

\begin{figure}[!t]
	\centering
	\includegraphics[width=2.4in]{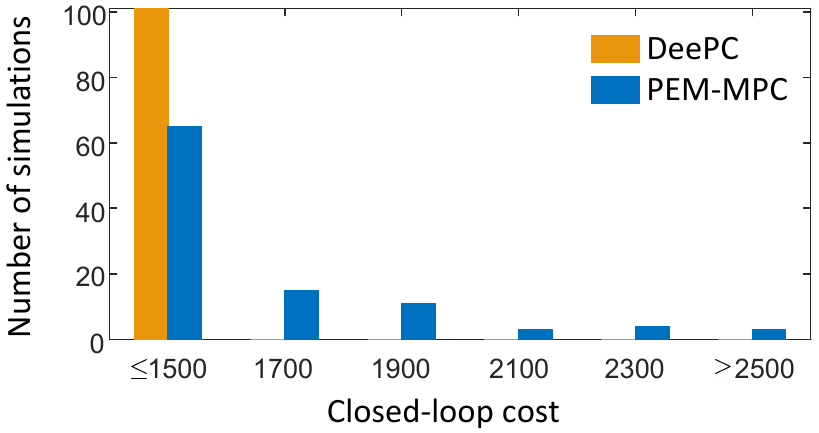}
	\vspace{-3mm}
	\caption{Cost comparison of DeePC and certainty-equivalence PEM-MPC in terms of closed-loop cost from $10{\rm s}$ to $30{\rm s}$ with $k = 60$.}
	\vspace{-1mm}
	\label{Fig_Linear_case_stat}
\end{figure}

The above simulations on DeePC and PEM-MPC were repeated $100$ times with different data sets to construct the Hankel matrices. The histogram in Fig.~\ref{Fig_Linear_case_stat} displays the closed-loop costs (i.e., $\sum_{i=500}^{1500} \|u_i\|_R^2 + \|y_i-r_i\|_Q^2$ measured from the system) from $10{\rm s}$ to $30{\rm s}$. It shows that DeePC consistently achieves superior closed-loop performance than certainty-equivalence PEM-MPC. This performance gap is due to the fact that PEM-MPC uses a nominal model (hence, certainty equivalence) without any robustification. {Of course, the nominal PEM-MPC can be further improved by considering robust identification and advanced MPC algorithms. However, we refrain from doing so to compare the basic DeePC to the basic PEM-MPC. Note that the performance of DeePC can also be improved with some algorithmic modifications.}

\begin{figure}[!t]
	\centering
	\includegraphics[width=2.8in]{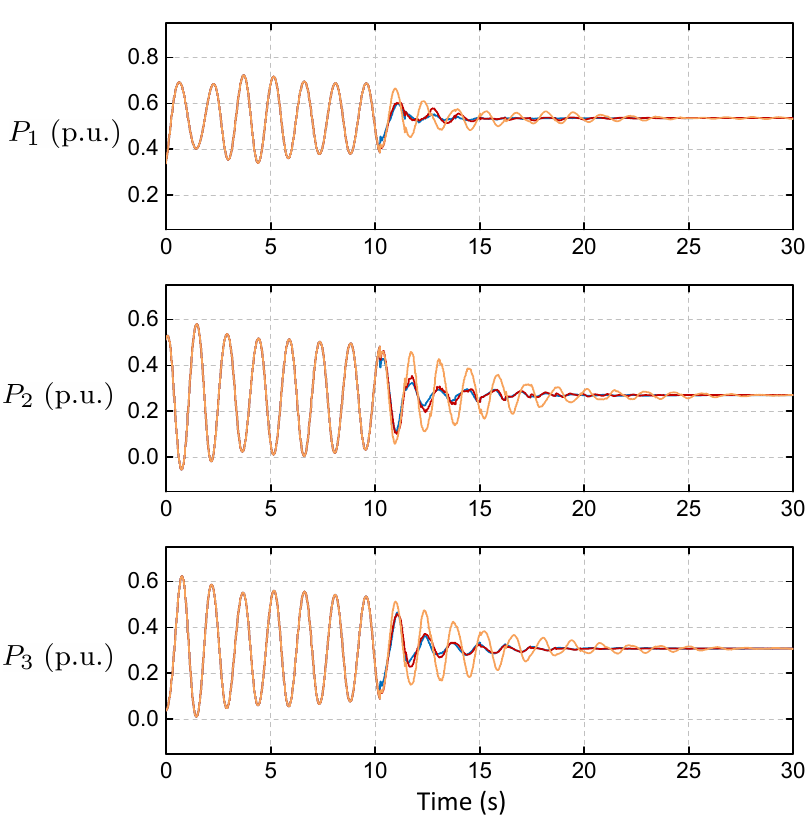}
	\vspace{-2mm}
	\caption{Comparison of performance when losing measurement signals. DeePC is activated at $t = 10{\rm s}$ ($k = 60$). {\color{CBLUE}{\bf{-----}}} No loss of measurement; {\color{CRED}{\bf{-----}}} Losing $P_2$; {\color{CYELLOW}{\bf{-----}}} Losing $P_2$ and $P_3$.}
	\vspace{-1mm}
	\label{Fig_loss_compare}
\end{figure}

{To further test the robustness of DeePC, we also implemented it in situations where some measurements become unavailable, due for example to sensing or communication failures. The controller was still able to stabilize the system even if $P_2$ or both $P_2$ and $P_3$ become unavailable (in the Hankel matrices and the initial trajectory, we eliminate the rows corresponding to
the lost measurement), with a modest reduction in performance, as shown in Fig.~\ref{Fig_loss_compare}.}

{The simulation results in this paper can be reproduced using the code available at: \url{https://www.research-collection.ethz.ch/handle/20.500.11850/469607}.}

\subsection{Nonlinear, Delayed and Noisy Implementation}
\label{subsec:nonlinear_DeePC}

To test the algorithms in a more practical setting of the four-area system we also considered the following conditions: a) the loads consist of constant power loads and nonlinear loads, e.g., induction motors (IMs) (here we use the same IM model and parameters as those in \cite{kawabe2014analytical}); b) load fluctuations are taken into account by adding white noise (noise power: $4\times 10^{-6}~{\rm p.u.}$) in the reference values of loads; c) the output measurements are noisy (noise power: $4\times 10^{-6}~{\rm p.u.}$); d) communication delays are considered (set as $100{\rm ms}$).

\begin{figure}[!t]
	\centering
	\includegraphics[width=2.8in]{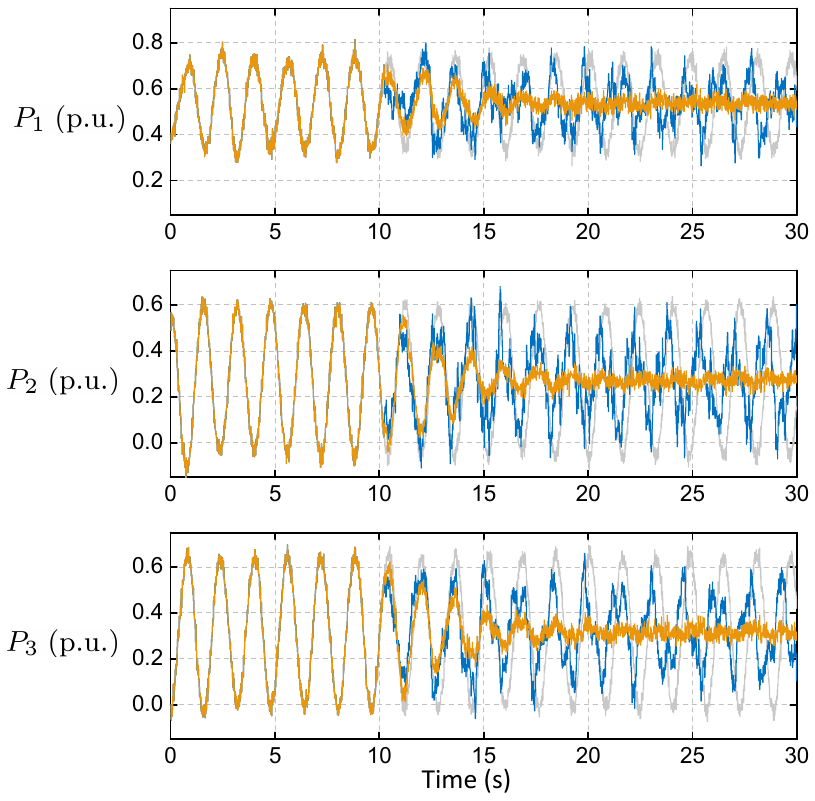}
	\vspace{-2mm}
	\caption{Time-domain responses of the four-area system with the practical setting. DeePC (or PEM-MPC) is activated at $t = 10{\rm s}$. {\color{CGRAY}{\bf{-----}}} without wide-area control; {\color{CBLUE}{\bf{-----}}} with PEM-MPC ($k = 60$); {\color{CYELLOW}{\bf{-----}}} with DeePC ($k = 60$).}
	\vspace{-0mm}
	\label{Fig_nonlinear_loads}
\end{figure}

Fig.~\ref{Fig_nonlinear_loads} shows the time-domain responses of the four-area system when the above settings are considered in the simulations. It can be seen that the low-frequency oscillations are mitigated with the DeePC algorithm. By comparison, the oscillations still exist when employing PEM-MPC. This is because DeePC does not rely on an explicit system model and therefore provides more flexibility than conventional MPC methods \cite{coulson2019data,huang2019data}.

\begin{figure}[!t]
	\centering
	\includegraphics[width=2.4in]{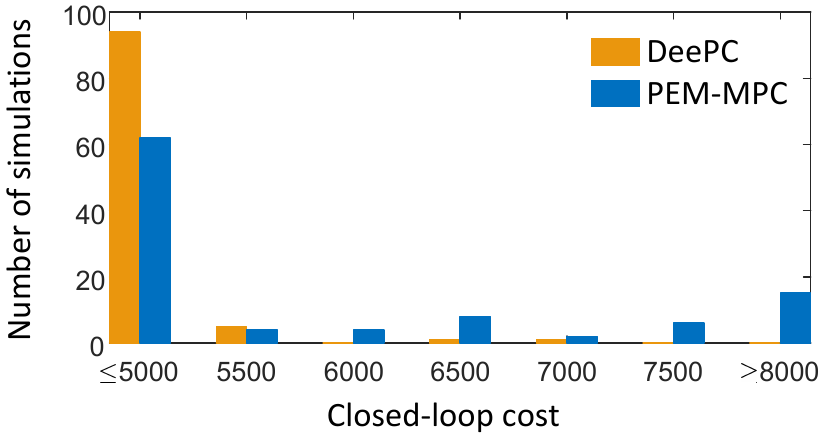}
	\vspace{-2mm}
	\caption{Closed-loop cost comparison of DeePC and certainty-equivalence PEM-MPC under the practical setting (from $10{\rm s}$ to $30{\rm s}$, $k = 60$).}
	\label{Fig_Nonlinear_case_stat}
\end{figure}

Repeating the simulations $100$ times with different data sets to construct the Hankel matrices and different random seeds for the measurement noise and load noise gives rise to the histogram in Fig.~\ref{Fig_Nonlinear_case_stat}. It is evident that DeePC achieves better performance than PEM-MPC on average. Moreover, the application of PEM-MPC may lead to instabilities of the system and thus unacceptable performance (e.g., with closed-loop performance larger than 8000 in Fig.~\ref{Fig_Nonlinear_case_stat}).

\subsection{DeePC Hyperparameter Tuning}

\begin{figure}[!t]
	\centering
	\includegraphics[width=3.3in]{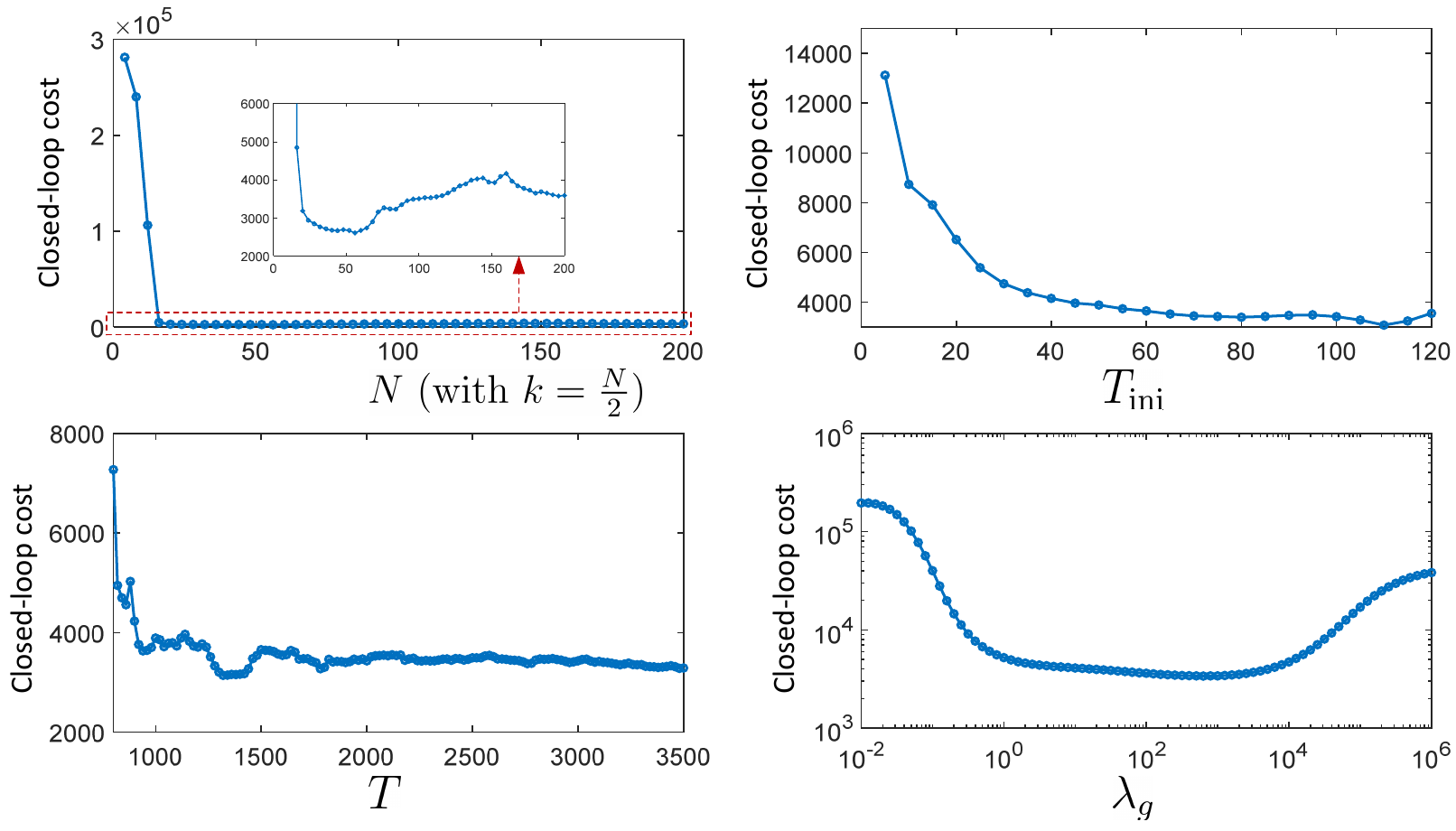}
	\vspace{-2mm}
	\caption{Closed-loop cost of the system with different DeePC parameters ($R = I$, $Q = 400 \times I$ and $\lambda_y = 2000$).}
	\vspace{-1mm}
	\label{Fig_N_T_Tini_lambda_g}
\end{figure}

We now discuss the parameter tuning of DeePC ($N$, $T_{\rm ini}$, $T$ and $\lambda_g$). Similar to conventional MPC, setting the prediction horizon $N$ large enough is required for stability. Fig.~\ref{Fig_N_T_Tini_lambda_g} plots the closed-loop cost (from $10{\rm s}$ to $30{\rm s}$) of the system with different parameters. The closed-loop cost dramatically drops with the increase of the prediction horizon $N$ and then remains within an acceptable range (in this plot we set $k=\frac{N}{2}$).

The initial trajectory determines the inherent system state, and thus $T_{\rm ini}$ gives a complexity for the model (related to the lag $\ell$ of the system). Fig.~\ref{Fig_N_T_Tini_lambda_g} shows that the closed-loop cost drops with the increase of $T_{\rm ini}$ from 5 to 40 and then remains nearly the same (as the system state is uniquely determined once $T_{\rm ini} \ge \ell$ in the deterministic case).


{According to \cite{markovskyidentifiability}, ${\rm rank}(\mathscr{H}_{T_{\rm ini}+N}(u,y)) = m(T_{\rm ini}+N)+n$ is required to make accurate prediction of the future behavior of the system. This, in turn, requires $T$ to be sufficiently large; note that, in particular, the size of the Hankel matrix grows linearly in the dimension, $n$, that we assume for the underlying system.}
Fig.~\ref{Fig_N_T_Tini_lambda_g} shows that the closed-loop cost significantly drops when $T$ is increased from 800 to 1000 and then remains nearly the same. We also observe that choosing a square Hankel matrix gives usually good performance, e.g., a minimum of the closed-loop cost (over $T$) appears in Fig.~\ref{Fig_N_T_Tini_lambda_g} around $T = 1439$ (corresponding to a square Hankel matrix), which indicates that incorporating more data may not necessarily provide better performance. We will explore this in future work.

As mentioned before, the regularization on $g$ in the cost function introduces distributional robustness \cite{coulson2019regularized}. Generally, the choice for $\lambda_g$ has a wide admissible range (relative to the choices of $R$ and $Q$). As displayed in Fig.~\ref{Fig_N_T_Tini_lambda_g}, the system has satisfactory performance for a wide range of $\lambda_g$.
{As shown in \cite{coulson2019regularized}, \cite{huang2020quadratic}, \cite{huang2021robust}, and \cite{coulson2020distributionally}, adding a regularization term on $g$ is equivalent to solving a (distributionally) robust optimization problem that minimizes the worst-case performance for an appropriately bounded disturbance set on the input/output data, where the bound is monotonically increasing in $\lambda_g$. Hence, if $\lambda_g$ is too small, the obtained optimal control sequence is not robust against uncertainties in the data, which could adversely impact closed-loop performance. If $\lambda_g$ is too large, the closed-loop performance could also be affected because the obtained optimal control sequence becomes conservative.}

In short, Fig.~\ref{Fig_N_T_Tini_lambda_g} indicates the robustness of the DeePC algorithm with regards to the choices of parameters. The system presents superior damping performance with proper regularization on $g$ and sufficiently large $N$, $T_{\rm ini}$ and $T$.

\subsection{Comparison of Hankel Matrix and Page Matrix}
\label{subsec: Comparison of Hankel Matrix and Page Matrix}

\begin{figure}[!t]
	\centering
	\includegraphics[width=3.5in]{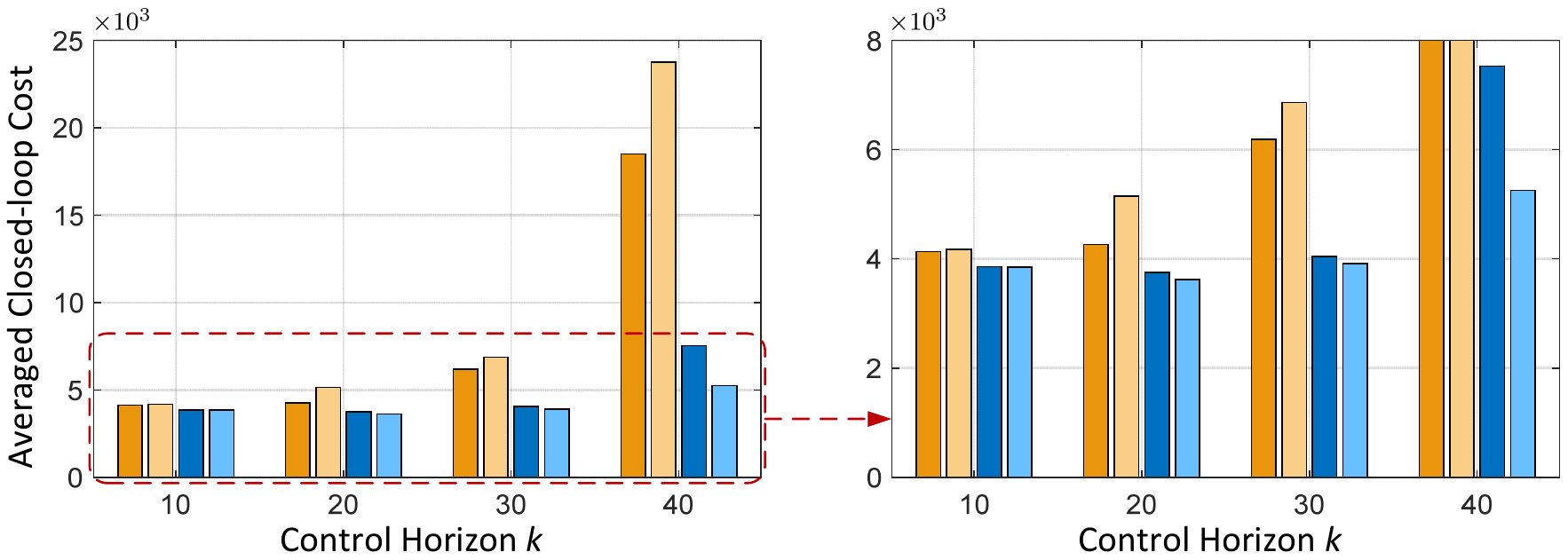}
	\vspace{-4mm}
	\caption{Averaged closed-loop cost when using Hankel matrix and Page matrix. \crule[CYELLOW]{0.25cm}{0.15cm} Hankel matrix without SVD; \crule[LYELLOW]{0.25cm}{0.15cm} Hankel matrix with SVD ($\sigma_0 = 1$); \crule[CBLUE]{0.25cm}{0.15cm} Page matrix without SVD; \crule[LBLUE]{0.25cm}{0.15cm} Page matrix with SVD ($\sigma_0 = 1$).}
	\vspace{-1mm}
	\label{Fig_Perf_compare}
\end{figure}

Fig.~\ref{Fig_Perf_compare} shows the averaged closed-loop cost of the system from $10{\rm s}$ to $30{\rm s}$ with different control horizon $k$ and different forms of data matrices (in the simulations, each case is repeated 100 times with different data sets to construct the Hankel/Page matrices and different random seeds for the measurement noise).
Here we choose a shorter prediction horizon ($N = 40$) to avoid an unacceptable value of $T$ to construct the Page matrices. We set $T = 1000$ in the simulations with Hankel matrices. To make sure that the Hankel matrices and the Page matrices have the same size, we set $T = 90100$ in the simulations with Page matrices; as expected, a much longer trajectory is required to construct the Page matrices. The other parameters are the same as those in Section~\ref{subsec:nonlinear_DeePC}.

It can be seen that using Page matrices in the DeePC algorithm achieves better performance than using Hankel matrices even without noise filtering based on SVD. We attribute this to the fact that the Page matrices are based on more data which thus contain more information about the system. The performance is further improved with noise filtering on the Page matrices, and the improvement is significant with a larger control horizon $k$. However, if we perform a similar noise-filtering process on the Hankel matrices, the performance deteriorates. We attribute this observation to the fact that the Hankel structure cannot be preserved after truncating some small singular values, as discussed in Section~\ref{subsec: DeePC with Page Matrix}.

\section{Min-Max DeePC}

The DeePC algorithm presented above acts as a centralized wide-area control, which is not resilient to communication failures, especially when more VSC-HVDC stations are considered. To alleviate this problem, we develop a Min-Max DeePC algorithm where inputs from a neighboring subsystem are modeled as disturbances in the spirit of Plug-and-play MPC or robust optimal control \cite{riverso2013plug,zhang2017robust,darivianakis2017power}. This enables a decentralized wide-area control implementation for oscillation damping, and is also useful to robustify DeePC against measured disturbances.

\subsection{Basic Formulation}\label{subsec_minmax_DeePC}

We extend the unknown LTI system in \eqref{eq:ABCD} by adding a measured disturbance vector $w_t \in \mathbb{R}^q$ to \eqref{eq:ABCD} as
\begin{equation}
\left\{ \begin{array}{l}
{x_{t + 1}} = A{x_t} + B{u_t} + E{w_t}\\
{y_t} = C{x_t} + D{u_t} + F{w_t}
\end{array} \right.\,,		\label{eq:ABCDEF}
\end{equation}
where $E \in \mathbb{R}^{n \times q}$ and $F \in \mathbb{R}^{p \times q}$.

To be specific, the unknown system is subjected to some external disturbances ($w_t$) whose past trajectory can be measured but whose future trajectory is unknown. Let $w^{\rm d}$ be a disturbance trajectory of length $T$ (i.e., $w^{\rm d} \in \mathbb{R}^{qT}$) measured from the unknown system such that ${\rm col}(u^{\rm d},w^{\rm d})$ is persistently exciting of order $T_{\rm ini}+N+n$. Note that here $w_t$ is regarded as an uncontrollable input vector of the unknown system. Similar to $u^{\rm d}$ and $y^{\rm d}$, we use $w^{\rm d}$ to construct the Hankel matrix  $\mathscr{H}_{T_{\rm ini}+N}(w^{\rm{d}})$, which is further partitioned into two parts as
\begin{equation}
\left[ {\begin{array}{*{20}{c}}
	{{W_{\rm P}}}\\
	{{W_{\rm F}}}
	\end{array}} \right] := \mathscr{H}_{T_{\rm ini}+N}(w^{\rm{d}})\,,		\label{eq:partition_Hw}
\end{equation}
where $W_{\rm P} \in \mathbb{R}^{qT_{\rm ini} \times H_c}$ and $W_{\rm F} \in \mathbb{R}^{qN \times H_c}$.
As in \eqref{eq:Hankel_g}, ${\rm col}(u_{\rm ini},w_{\rm ini},y_{\rm ini},u,w,y)$ is then a trajectory of the unknown system \eqref{eq:ABCDEF} if and only if there exists $g \in \mathbb{R}^{H_c}$ so that

\begin{equation}
\left[ {\begin{array}{*{20}{c}}
	{{U_{\rm P}}}\\
	{{W_{\rm P}}}\\
	{{Y_{\rm P}}}\\
	{{U_{\rm F}}}\\
	{{W_{\rm F}}}\\
	{{Y_{\rm F}}}
	\end{array}} \right]g = \left[ {\begin{array}{*{20}{c}}
	{{u_{\rm ini}}}\\
	{{w_{\rm ini}}}\\
	{{y_{\rm ini}}}\\
	u\\
	w\\
	y
	\end{array}} \right]\,,		\label{eq:Hankel_g_w}
\end{equation}
where $w_{\rm ini} \in \mathbb{R}^{qT_{\rm ini}}$ is the most recent measured disturbance trajectory and $w = {\rm col}(w_0,w_1,...,w_{N-1}) \in \mathbb{R}^{qN}$ is the future disturbance trajectory. We assume that this future trajectory is unknown but bounded with $w_t \in \left[\underline w, \overline w\right]^q$.

The Min-Max DeePC algorithm solves the following robust optimization problem
\begin{equation}
\begin{array}{l}
\mathop {{\rm{min}}}\limits_{g,\sigma_y,u \in \mathcal U, y \in \mathcal Y} \; \mathop {{\rm{max}}}\limits_{w \in \mathcal W}\;\;{\left\| u \right\|_R^2} + {\left\| {y - r} \right\|_Q^2} + {\lambda _g}{\left\| g \right\|_2^2}  + {\lambda _y}{\left\| \sigma_y \right\|_2^2}\\
{\rm s.t.}\;\;\left[ {\begin{array}{*{20}{c}}
	{{U_{\rm P}}}\\
	{{W_{\rm P}}}\\
	{{Y_{\rm P}}}\\
	{{U_{\rm F}}}\\
	{{W_{\rm F}}}\\
	{{Y_{\rm F}}}
	\end{array}} \right]g = \left[ {\begin{array}{*{20}{c}}
	{{u_{\rm ini}}}\\
	{{w_{\rm ini}}}\\
	{{y_{\rm ini}}}\\
	u\\
	w\\
	y
	\end{array}} \right] + \left[ {\begin{array}{*{20}{c}}
	0\\
	0\\
	\sigma_y\\
	0\\
	0\\
	0
	\end{array}} \right]\,,		
\label{eq:MinMax_DeePC}
\end{array}
\end{equation}
where $\mathcal W = \left[\underline w,\overline w\right]^{qN} \subseteq \mathbb{R}^{qN}$ is the disturbance constraint set imposing upper and lower bounds on $w_t$. One may also consider a slack variable $\sigma_w$ added to $w_{\rm ini}$ under noisy measurements. Similar to the DeePC algorithm, \eqref{eq:MinMax_DeePC} is implemented in a receding horizon fashion.
By solving the robust optimization problem in \eqref{eq:MinMax_DeePC}, the Min-Max DeePC provides robust and optimal control inputs with regards to the worst case of the future disturbance trajectory within the set $\mathcal W$.

Next, we will show how to remove the equality constraints so that \eqref{eq:MinMax_DeePC} can be solved by standard robust optimization solvers.
Let $H = {\rm col}(U_{\rm P},W_{\rm P},Y_{\rm P},U_{\rm F},W_{\rm F})$ and $x_{\rm ini} = {\rm col}(u_{\rm ini},w_{\rm ini},y_{\rm ini}+\sigma_y,u,w)$ such that $Hg=x_{\rm ini}$. Then, the solution of $Hg=x_{\rm ini}$ can be obtained by
\begin{equation}
g = H^+x_{\rm ini} + H^\bot x\,,
\label{eq:g_x}
\end{equation}
where $H^\bot = I - H^+H$ ($I$ is the identity matrix), and $x$ can be any vector in $\mathbb{R}^{H_c}$. Further, we have
\begin{equation}
y = Y_{\rm F}g = Y_{\rm F}H^+x_{\rm ini} + Y_{\rm F}H^\bot x\,.
\label{eq:y_x}
\end{equation}

{By substituting \eqref{eq:g_x} and \eqref{eq:y_x} into the objective function of \eqref{eq:MinMax_DeePC} we eliminate the decision variables $g,y$ and thus the equality constraints. Then, we reformulate the optimization problem in its epigraph form and derive the robust counterpart so that it becomes a conic program that can be solved by standard solvers \cite{lofberg2012automatic}.} {Note that by eliminating $g$ and $y$ in \eqref{eq:MinMax_DeePC}, we implicitly assume that $g$ and $y$ are (second-stage) adjustable decision variables that are decided after the disturbances are revealed \cite{zhen2018adjustable}. Moreover, by applying \eqref{eq:g_x} and \eqref{eq:y_x} we assume that $g$ and $y$ are linear functions of $w$, and the obtained formulation is a relaxation of \eqref{eq:MinMax_DeePC} as equality constraints in a min-max problem are in general conservative and may be infeasible (hence relaxation is needed). We will consider different relaxation or approximation methods for \eqref{eq:MinMax_DeePC} in future work.}


\begin{figure*}[!t]
	\centering
	\includegraphics[width=5in]{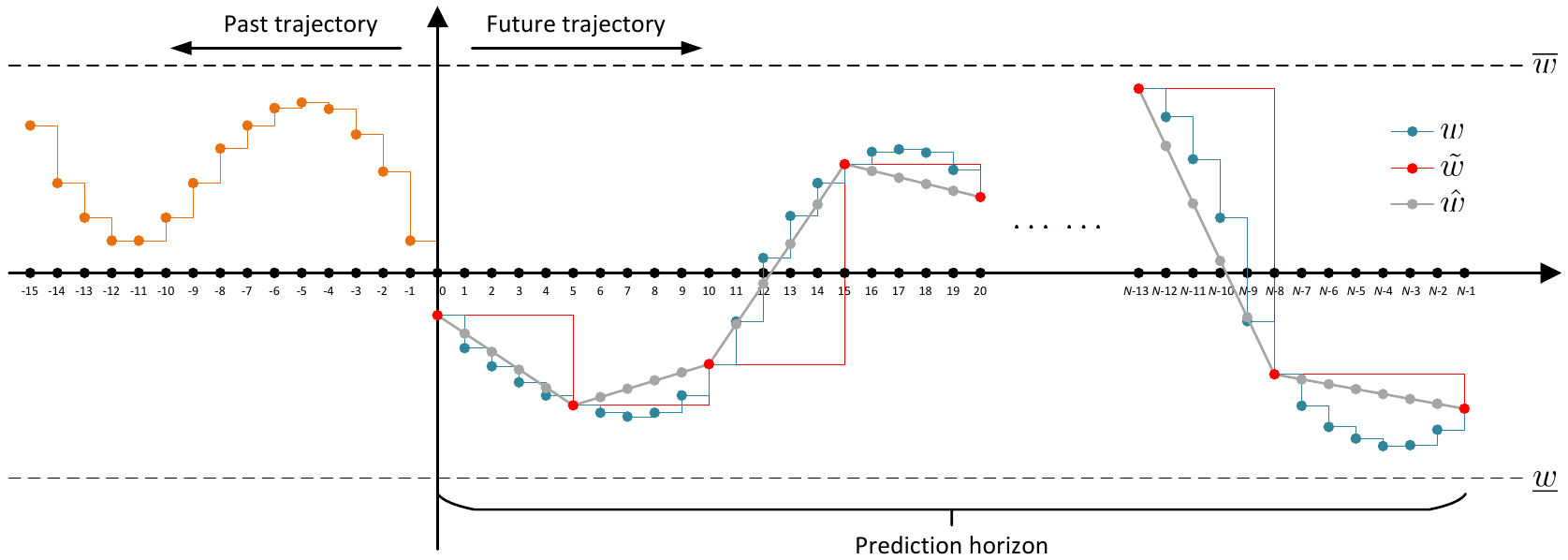}
	\vspace{-1mm}
	\caption{Downsampling of the future disturbance trajectory $w$.}
	\vspace{-1mm}
	\label{Fig_resampling}
\end{figure*}

\subsection{Disturbance-Feedback (DF) Min-Max DeePC}

Similar to conventional Min-Max MPC, the Min-Max DeePC algorithm could be unnecessarily conservative, as it ignores the feedback (recourse) implicit in the receding horizon implementation. The control sequence obtained by solving \eqref{eq:MinMax_DeePC} is optimal in an open-loop sense. However, as the control horizon is typically shorter than the prediction horizon, feedback is introduced every time \eqref{eq:MinMax_DeePC} is re-solved, by measuring results of previous control actions and disturbances and accordingly updating the future control sequence. This feedback is not transparent and actually ignored in \eqref{eq:MinMax_DeePC} leading to potentially conservative control sequences.

Closed-loop Min-Max MPC approaches have been developed to reduce this conservativeness. They assume that the input at every time would be calculated with the knowledge of the current system state \cite{scokaert1998min,lee1997worst,lofberg2003approximations}. For example, (approximate) dynamic programming can be used to optimize over a general class of feedback policies. This approach, however, comes with very high computational cost and can only be applied to small systems with short horizons. Alternatively, one can parameterize the dependence of the control decisions on the state and/or disturbance using a more limited class of functions, which emulates the effects of feedback in the receding-horizon implementation \cite{lofberg2003approximations,bemporad1998reducing}.
Inspired by \cite{lofberg2003approximations} and \cite{goulart2006optimization}, we apply the following affine DF policy
\begin{equation}\label{eq:Lw}
u = v + {\mathcal L}w,
\end{equation}
where $v \in \mathbb{R}^{mN}$ is a new control variable, and we assume that the feedback matrix $\mathcal L \in \mathbb R ^ {mN \times qN}$ has a strictly lower block triangular Toeplitz structure to enforce causality and reduce complexity of $\mathcal L$ to $mq(N-1)$ independent entries \cite{lofberg2003approximations}.

Then, we introduce a DF Min-Max DeePC algorithm which solves the following robust optimization problem
\begin{equation}
\begin{array}{l}
\hspace{-2.5mm}\mathop {{\rm{min}}}\limits_{\scriptsize \begin{array}{c} g,\sigma_y, \mathcal L, y \in \mathcal Y, \\v \in \mathcal U \end{array}} \hspace{-0.7mm}\mathop {{\rm{max}}}\limits_{w \in \mathcal W}\;\;
{\left\| v \right\|_R^2} + {\left\| {y - r} \right\|_Q^2} + {\lambda _g}{\left\| g \right\|_2^2} + {\lambda _y}{\left\| \sigma_y \right\|_2^2}  \\
{\rm s.t.}\;\;\left[ {\begin{array}{*{20}{c}}
	{{U_{\rm P}}}\\
	{{W_{\rm P}}}\\
	{{Y_{\rm P}}}\\
	{{U_{\rm F}}}\\
	{{W_{\rm F}}}\\
	{{Y_{\rm F}}}
	\end{array}} \right]g = \left[ {\begin{array}{*{20}{c}}
	{{u_{\rm ini}}}\\
	{{w_{\rm ini}}}\\
	{{y_{\rm ini}}}\\
	v + {\mathcal L}w\\	w\\	y	\end{array}} \right] + \left[ {\begin{array}{*{20}{c}}
	0\\	0\\	\sigma_y\\	0\\	0\\	0	\end{array}} \right]\,,
\label{eq:MinMax_DeePC_Lw}
\end{array}
\end{equation}
where $v$ and $\mathcal L$ in the DF policy are now decision variables to be optimized over. We remark that the disturbance feedback term $\mathcal L w$ is included in the above robust optimization problem to implicitly emulate the effects of feedback, or to be more specific, the updates of $w_{\rm ini}$ in the receding-horizon implementation. After solving \eqref{eq:MinMax_DeePC_Lw}, the first $k$ elements of $v$ will be applied to the system.

To solve the robust optimization problem in \eqref{eq:MinMax_DeePC_Lw}, we eliminate the equality constraints and rewrite it in epigraph form, similar to the process in Section~\ref{subsec_minmax_DeePC}. Notice that the bilinear term $\mathcal L w$ in \eqref{eq:MinMax_DeePC_Lw} makes the robust optimization problem difficult to solve. Fortunately, the difficulty can be eased  by using a semidefinite relaxation transforming the epigraph constraint into a matrix inequality, as detailed in \cite{lofberg2003approximations} and \cite{lofberg2003minimax}. To reduce the computational burden, here we ignore the regularization of $g$ (by setting $\lambda_g = 0$) in the cost function such that the resulting matrix inequality has a lower dimension.

\subsection{Downsampling of Future Disturbance Trajectory}

Our parameterization of the future disturbance trajectory $w \in \mathcal W \subseteq \mathbb{R}^{qN}$ can be of high dimension when we choose a long prediction horizon, leading to a high computational burden when solving the robust optimization problem in \eqref{eq:MinMax_DeePC} and \eqref{eq:MinMax_DeePC_Lw}. We discuss how to relieve the computational burden by constraining the set $\mathcal W$ and thus reducing the dimension of the future disturbance trajectory.

Notice that normally disturbances are not random bounded signals but have a certain degree of smoothness especially when low-frequency dynamics are considered. In fact, exploiting the correlation existing in disturbances is an efficient way to reduce the uncertainty \cite{zhang2017robust,darivianakis2017power}. In what follows, we show how to bound the bandwidth or total variation of the disturbance.
In a first step we perform downsampling on $w$ by selecting one every $M$ steps of $w$ to get the lower-dimensional representation $\tilde w \in \mathbb{R}^{q[{\mathscr R}(N/M)+1]}$ (the function ${\mathscr R}(a)$ rounds $a$ to the nearest integer toward zero).

As shown in Fig.~\ref{Fig_resampling}, the downsamping leads to a lower-dimensional, but less accurate representation of the future disturbance trajectory. To smoothen this low-dimension trajectory and bring it to the same sampling rate as $u$ and $y$, we linearly interpolate on $\tilde w$, leading to an extended trajectory $\hat w$ (illustrated in Fig.~\ref{Fig_resampling}) given by
\begin{equation}
\hat w_i = \left\{ \begin{array}{l}
\begin{split}
\tilde w_{{\mathscr R}(i/M)} + (i \; {\rm mod} \; M) \times \frac{\tilde w_{{\mathscr R}(i/M)+1} - \tilde w_{{\mathscr R}(i/M)}}{M},&\\\;0 \le i \le \bar i,&\end{split}\\
\begin{split}
\tilde w_{{\mathscr R}(N/M)-1} + (i - \bar i) \frac{\tilde w_{{\mathscr R}(N/M)} - \tilde w_{{\mathscr R}(N/M)-1}}{N - 1 - \bar i},&\\\;\bar i < i \le N-1\,.&\end{split}
\end{array} \right.
\label{eq:upsampling}
\end{equation}
where $\bar i = M[{\mathscr R}(N/M) - 1]$, and $A \; {\rm mod} \; B$ denotes the remainder of $\frac{A}{B}$.

By replacing $w$ by $\hat w$ in \eqref{eq:MinMax_DeePC} and \eqref{eq:MinMax_DeePC_Lw} we obtain a modified version of the Min-Max DeePC algorithms which have lower-dimensional uncertainty parameterization because $\hat w$ entirely depends on $\tilde w$, thereby leading to lower computational burden.
We note that the signal space of $\hat w$ is in fact a subspace of that of $w$, that is, by maximizing over $\hat w$ one may not include the worst case in \eqref{eq:MinMax_DeePC} and \eqref{eq:MinMax_DeePC_Lw} unless the disturbance signal is itself smooth and satisifies \eqref{eq:upsampling}. In the next section, we will show that by imposing \eqref{eq:upsampling} we can in fact get satisfactory performance when dealing with low-frequency oscillations.

\section{Decentralized Wide-Area Control}

We now apply the Min-Max DeePC algorithm in the four-area test system to perform decentralized, robust, and optimal wide-area control (the parameters of the four-area system are the same as those in Section~\ref{subsec:nonlinear_DeePC}).
In a first step, the four-area system is partitioned into two (two-area) subsystems which both receive two external inputs ($P_3$ and $P_{\rm dc}$) as shown by the dashed red lines in Fig.~\ref{Fig_four_area_syst}. The past trajectories of $P_3$ and $P_{\rm dc}$ are measured, but their future trajectories are unpredictable from the subsystem point of view.

\begin{figure}[!t]
	\centering
	\includegraphics[width=3in]{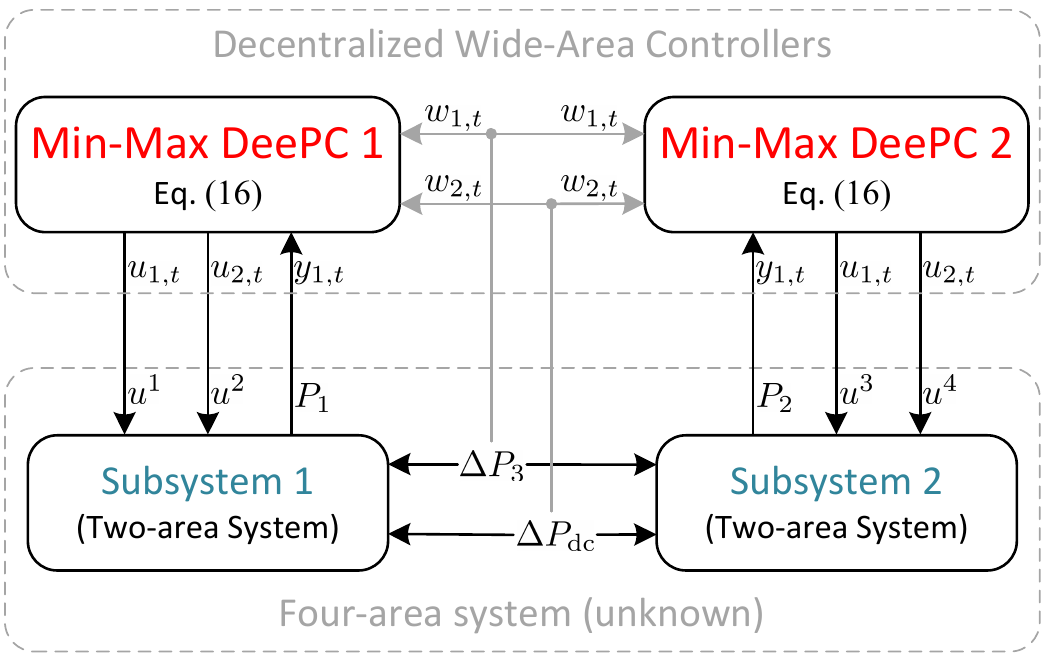}
	\vspace{-4mm}
	\caption{Decentralized wide-area control based on Min-Max DeePC.}
	\vspace{-2mm}
	\label{Fig_Min_Max_DeePC}
\end{figure}

Each subsystem employs a wide-area controller to provide safe and robust optimal control policies obtained from \eqref{eq:MinMax_DeePC} for the VSC-HVDC station within it, denoted by \text{\em Min-Max DeePC 1} and \text{\em Min-Max DeePC 2} in Fig.~\ref{Fig_Min_Max_DeePC}.
We choose $P_1$ from Subsystem 1 as the output signal for Min-Max DeePC 1 such that VSC-HVDC station 1 aims at mitigating the oscillation in $P_1$; the symmetric holds for Subsystem 2.
The deviations of the signals $P_3$ and $P_{\rm dc}$ from their steady-state values are considered as the external disturbances (i.e., $w_{1,t}=\Delta P_3$ and $w_{2,t}=\Delta P_{\rm dc}$) in the Min-Max DeePC algorithms, that is, Min-Max DeePC 1 and Min-Max DeePC 2 provide robust optimal control policies over the worst future trajectories that may occur in $P_3$ and $P_{\rm dc}$. Under the above setting, every controller needs one local measurement ($P_{\rm dc}$) and two wide-area measurements ($P_3$ and $P_1$, or $P_2$).

Since each subsystem is about half of the size of the original system, we choose a smaller $T_{\rm ini}=30$. The prediction horizon is chosen to be $N = 40$ (i.e., we predict forward $0.8{\rm s}$) to reduce the number of the decision variables and thus the computational burden.
The reduction factor $M$ of $w$ is set to $40$ to reduce the dimension of uncertainties, that is, only the first and last points of the disturbance trajectories are considered as uncertain and the other points in between are obtained by linear interpolation. The upper and lower bounds for $w$ are set to $\overline w = 0.3$ and $\underline w = -0.3$. Note that we focus only on the low-frequency oscillations in $w$ which justifies the downsampling approach. Moreover, we force  $x$ in \eqref{eq:g_x} and \eqref{eq:y_x} to be zero to reduce the number of decision variables, which will in fact lead to a suboptimal solution for the Min-Max DeePC if the system is not LTI or noise-free. 
The coefficients in the cost function are the same as those in Section~\ref{subsec:Centralized Wide-area Control}.
Before activating the Min-Max DeePC in each VSC-HVDC station, persistently exciting white noise signals (noise power: $10^{-4}~{\rm p.u.}$) are injected into the system (through $u^1$, $u^2$, $u^3$ and $u^4$) for $10{\rm s}$ (with $T = 500$) to get the data Hankel matrices \eqref{eq:partition_Huy} and \eqref{eq:partition_Hw}.

\begin{figure}[!t]
	\centering
	\includegraphics[width=2.8in]{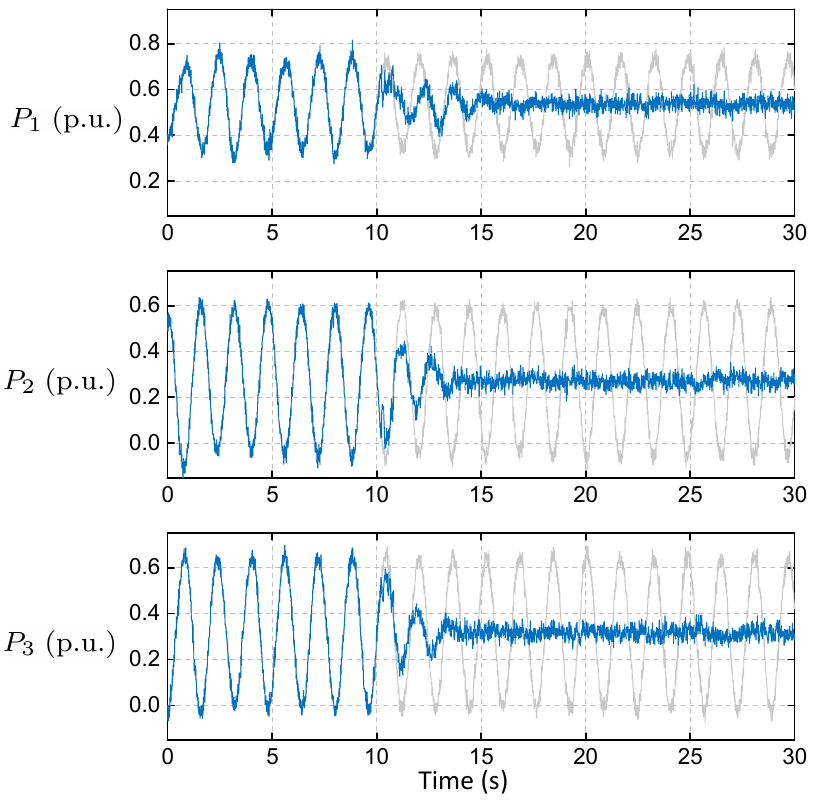}
	\vspace{-2mm}
	\caption{Time-domain responses of the four-area system with Min-Max DeePC. The Min-Max DeePC is activated at $t = 10{\rm s}$. {\color{CGRAY}{\bf{-----}}} without wide-area control; {\color{CBLUE}{\bf{-----}}} with Min-Max DeePC ($k = 8$).}
	\vspace{0mm}
	\label{Fig_Min_Max_cases}
\end{figure}

Fig.~\ref{Fig_Min_Max_cases} plots the time-domain responses of the four-area system with application of the Min-Max DeePC algorithm in \eqref{eq:MinMax_DeePC} mitigating the inter-area oscillations.
Here we use the YALMIP toolbox to solve the robust optimization problem in \eqref{eq:MinMax_DeePC} \cite{Lofberg2004,lofberg2012automatic}, with Mosek set as the solver for conic programs \cite{mosek2015mosek}. {Under the above configuration, the dimension of the decision variables in the conic program is $mN+pT_{\rm ini}$, the dimension of the uncertain variables is $2q$, and the number of inequality constraints is $2(m+p)N$.} It takes about $0.14{\rm s}$ to solve the robust optimization problem on an Intel Core i5 7200U CPU with 8GB RAM. Therefore, with this set-up, the sampling time of 0.02s, and by choosing $k$ no less than $8$, the Min-Max DeePC can be solved in real time.

\begin{figure}[!t]
	\centering
	\includegraphics[width=2.8in]{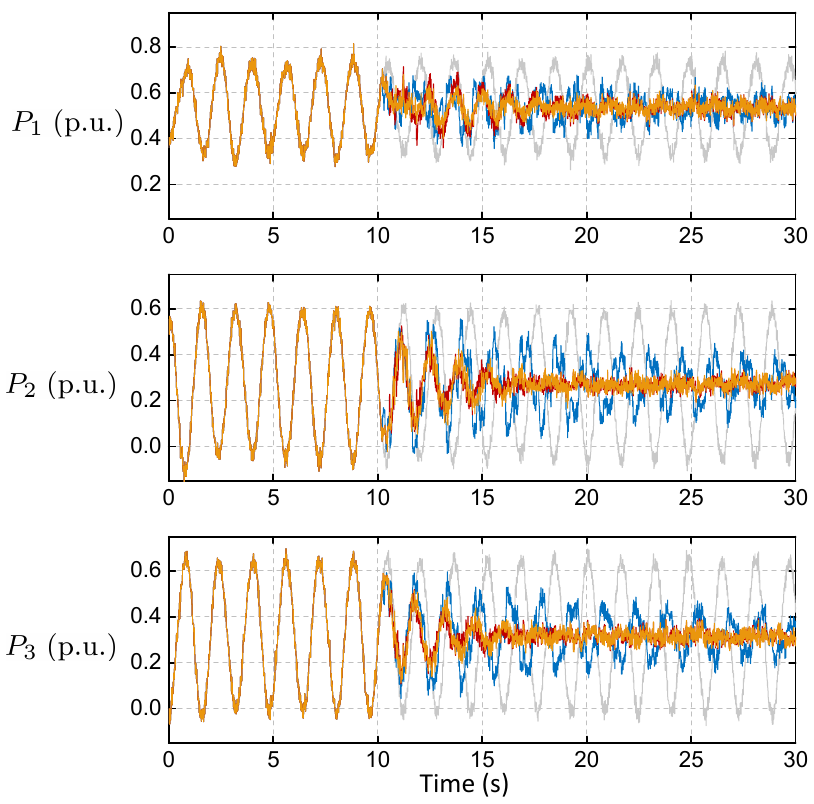}
	\vspace{-2mm}
	\caption{Time-domain responses of the four-area system with DF Min-Max DeePC. The algorithm is activated at $t = 10{\rm s}$. {\color{CGRAY}{\bf{-----}}} without wide-area control; {\color{CBLUE}{\bf{-----}}} with Min-Max DeePC ($k = 16$); {\color{CRED}{\bf{-----}}} with DF Min-Max DeePC ($k = 16$); {\color{CYELLOW}{\bf{-----}}} with DF Min-Max DeePC ($k = 24$);}
	\vspace{-2mm}
	\label{Fig_Min_Max_cases_Lw}
\end{figure}

Fig.~\ref{Fig_Min_Max_cases_Lw} shows the time-domain responses of the system with a comparison on the damping performance of Min-Max DeePC and DF Min-Max DeePC. We choose a longer control horizon ($k = 16$) to see how the algorithms perform when fast feedback is not available. It can be seen that the damping ratio of Min-Max DeePC with $k=16$ is significantly lower than with $k=8$ shown  in Fig.~\ref{Fig_Min_Max_cases}. By comparison, the DF Min-Max DeePC eliminates the oscillations with a much higher damping ratio, which we attribute to the reduced conservativeness of DF Min-Max DeePC. Moreover, even with $k=24$, the damping ratio remains almost the same when using DF Min-Max DeePC; under this setting the Min-Max DeePC algorithm would not be able to eliminate the oscillations. In short, the DF policy allows us to employ longer horizons, which again translates to more time for solving the optimization. We again use the YALMIP toolbox to solve the robust optimization problem, and the solving time is about 0.44s on an Intel Core i5 7200U CPU with 8GB RAM. Hence, by setting $k \ge 22$, the DF Min-Max DeePC algorithm can be implemented in real time.

\section{Conclusions}

We applied the DeePC algorithm as a model-free approach to perform optimal wide-area control based solely on input/output trajectories measured from the unknown system to predict the future behaviours. In the power systems context, DeePC utilizes the high controllability and flexibility of VSC-HVDC stations to mitigate low-frequency oscillations. We showed that even with nonlinear loads, load fluctuations, communication delays and noisy measurements, DeePC still effectively attenuates the inter-area oscillations in the system. We showed that by using Page matrices together with noise filtering based on SVD, the DeePC algorithm can achieve significantly better performance. Furthermore, we presented a Min-Max DeePC algorithm to enable decentralized, robust, and optimal wide-area control and discussed how to relieve the computational burden through downsampling of the future disturbance trajectory. Then, a disturbance feedback policy was introduced to reduce the conservativeness by considering the effects of feedback when solving the robust optimization problem. We showcased that the decentralized Min-Max DeePC algorithm effectively mitigates the inter-area oscillations and improves the scalability and reliability of the optimal wide-area control since a centralized controller is not needed. {We will develop adaptive DeePC algorithms to deal with time-varying dynamics (and in particular time varying delays) in future work.}

\section*{Acknowledgment}

The authors would like to thank Jianzhe Zhen for fruitful discussions, and Johan L{\"o}fberg for his useful suggestions on coding in YALMIP.


\appendices
\vspace{0mm}
\section{System parameters}
\label{Appendix: System parameters}
See Table~\ref{table:sys_parameters}.

\renewcommand{\thetable}{\thesection.\arabic{table}}
\setcounter{table}{0}
\renewcommand\arraystretch{1.25}
\begin{table}[h]
	\scriptsize
	\centering
	\vspace{-4mm}
	\caption{Parameters of the four-area test system}
	\begin{tabular}{|llll|}
		\hline
		\multicolumn{4}{|c|}{\bf Main parameters of the VSC-HVDC link (per-unit values)}				\\
		\hline
		\multicolumn{2}{|l}{Converter-side inductors: $L = 0.05$}			
		&\multicolumn{2}{l|}{\textit{LCL} capacitors:	$C_{\rm F} = 0.05$}								\\
		\multicolumn{2}{|l}{Grid-side inductors: $L_{\rm g} = 0.05$}			
		&\multicolumn{2}{l|}{Grid-side resistors: $R_{\rm g} = 0.01$}									\\
		\multicolumn{2}{|l}{DC-side capacitors: $C_{\rm dc} = 0.06$}			
		&\multicolumn{2}{l|}{DC-link resistors: $R_{\rm dc} = 0.015$}									\\
		\multicolumn{4}{|l|}{PI gains of the PLL: $103.1({\rm{rad/s}}),5311.5({\rm{rad/s}})$}			\\
		\multicolumn{4}{|l|}{PI gains of the current control loop: $0.3({\rm{p.u.}}),10({\rm{p.u.}})$}	\\
		\multicolumn{4}{|l|}{PI gains of the voltage control loop: $4({\rm{p.u.}}),40({\rm{p.u.}})$}	\\
		\multicolumn{4}{|l|}{PI gains of the power control loop: $0.2({\rm{p.u.}}),2({\rm{p.u.}})$}		\\
		\multicolumn{4}{|l|}{PI gains of the dc voltage control loop: $5({\rm{p.u.}}),50({\rm{p.u.}})$}	\\
		\hline
		\multicolumn{4}{|c|}{\bf Main parameters of the SGs (per-unit values)}							\\
		\hline
		$X_d = 2.065$		&	$X_q = 1.974$		&	$X'_d = 0.4879$			&	$X'_q = 1.19$		\\
		$X''_d = 0.35$		&	$X''_q = 0.35$		&	$T'_{d0} = 6.56$		&	$T'_{q0} = 1.5$		\\
		$T''_{d0} = 0.05$	&	$T''_{q0} = 0.035$	&	$J_{\rm SG} = 8.658$	&	$R_a = 0.0025$		\\
		\hline
		\multicolumn{4}{|c|}{Fast exciters (IEEET1 Model)}												\\
		\hline
		$K_A = 50$			&	$T_A = 0.05$		&	$K_F = 0.0057$			&	$T_F = 0.5$			\\
		$T_R = 0.1$			&						&							&						\\
		\hline
		\multicolumn{4}{|c|}{Steam Turbine and Governor (IEEEG1 Model)}									\\
		\hline
		$T_1 = 0.5$			&	$T_2 = 1$			&	$T_3 = 0.6$				&	$T_4 = 0.6$			\\
		$T_5 = 0.5$			&	$T_6 = 0.8$			&	$T_7 = 1$				&	$K = 5$				\\
		$K_1 = 0.3$			&	$K_2 = 0$			&	$K_3 = 0.25$			&	$K_4 = 0$			\\
		$K_5 = 0.3$			&	$K_6 = 0$			&	$K_7 = 0.15$			&	$K_8 = 0$			\\
		\hline
		\multicolumn{4}{|c|}{\bf Impedance of lines and power consumption of loads (per-unit values)}	\\
		\hline
		\multicolumn{2}{|l}{Line 1-5 \& 11-15: $0.005+j0.05$}			
		&\multicolumn{2}{l|}{Line 2-5 \& 12-15: $0.02+j0.2$}											\\
		\multicolumn{2}{|l}{Line 5-6 \& 15-16: $0.002+j0.02$}			
		&\multicolumn{2}{l|}{Line 6-10 \& 12-20: $0.004+j0.04$}											\\
		\multicolumn{2}{|l}{Line 6-7 \& 16-17: $0.01+j0.2$}			
		&\multicolumn{2}{l|}{Line 7-8 \& 17-18: $0.014+j0.28$}											\\
		\multicolumn{2}{|l}{Line 8-9 \& 18-19: $0.004+j0.08$}			
		&\multicolumn{2}{l|}{Line 8-18: $0.012+j0.12$}													\\
		\multicolumn{2}{|l}{Line 9-3 \& 19-13: $0.05+j0.02$}			
		&\multicolumn{2}{l|}{Line 9-4 \& 19-14: $0.05+j0.15$}											\\
		\multicolumn{2}{|l}{$P_{\rm Load1} = 0.9493$ (IM: 0.5)}			
		&\multicolumn{2}{l|}{$P_{\rm Load2} = 1.3$ (IM: 1.2)}											\\
		\multicolumn{2}{|l}{$P_{\rm Load3} = 0.7$ (IM: 0.2)}			
		&\multicolumn{2}{l|}{$P_{\rm Load4} = 1.7$ (IM: 1.4)}											\\
		\multicolumn{2}{|l}{$C_1$ \& $C_3$: $0.25$}			
		&\multicolumn{2}{l|}{$C_2$ \& $C_4$: $0.15$}													\\
		\hline
	\end{tabular}		
	\label{table:sys_parameters}
\end{table}

\bibliographystyle{IEEEtran}

\bibliography{ref}

\end{document}